\documentclass[longauth]{aa} 
\usepackage{sidecap}
\usepackage{xcolor}
\usepackage[colorlinks=true, allcolors=cyan]{hyperref}
\usepackage{subfigure}
\usepackage{lscape}
\usepackage{graphicx}
\usepackage{float} 
\usepackage{txfonts}
\newcommand{\hi}{{\rm H}\,{\rm \small I}}

\newcommand{\kms}{\,km\,s$^{-1}$}
\newcommand{\msun}{M$_{\odot}$}

\begin{document}

   \title{The molecular gas of four ram pressure stripped galaxies in the core of the Hydra I cluster}

  \author{
  Clara C. de la Casa\inst{\ref{iaa},\ref{eso}},
  Kelley M.~Hess\inst{\ref{chalmers},\ref{astron}},
  Nikki Zabel \inst{\ref{uct},\ref{saao}},
  Emanuela Pompei \inst{\ref{eso}},
  Yara L. Jaffé \inst{\ref{ufsm},\ref{mingal}},
  Nazir Makda \inst{\ref{saaos}},
  Julia Healy \inst{\ref{uct},\ref{jodrell},\ref{ukskao}},
  Johanna Hartke \inst{\ref{finca},\ref{tobs},\ref{tcs}},
  Lourdes Verdes-Montenegro\inst{\ref{iaa}},
  Enrichetta Iodice \inst{\ref{INAF}},
  Sarah Blyth \inst{\ref{saaos}},
  Ralf Kotulla\inst{\ref{uwisc}},
  Marie-Lou Gendron-Marsolais\inst{\ref{laval}},
  Renée C. Kraan-Korteweg\inst{\ref{uct}},
  Brenda Namumba\inst{\ref{iaa}},
  Amidou Sorgho\inst{\ref{iaa}},
  Roger Ianjamasimanana \inst{\ref{iaa}},
  \and
  John S. Gallagher\inst{\ref{wsu},\ref{maca}}}
  
    \authorrunning{de la Casa, C.~C.~et al.}
    
  \institute{
    Instituto de Astrof\'{i}sica de Andaluc\'{i}a (CSIC), Glorieta de la Astronom\'{i}a s/n, 18008 Granada, Spain \label{iaa}
    \and European Southern Observatory, Alonso de Córdova 3107, Vitacura, Santiago, Chile \label{eso}
    \and Department of Space, Earth and Environment, Chalmers University of Technology, Onsala Space Observatory, 43992 Onsala, Sweden \label{chalmers}
    \and ASTRON, the Netherlands Institute for Radio Astronomy, Postbus 2, 7990 AA, Dwingeloo, The Netherlands \label{astron}
    \and Department of Astronomy, University of Cape Town, Private Bag X3, 7701 Rondebosch, South Africa \label{uct}
    \and South African Astronomical Observatory, PO Box 9, Observatory Cape Town 7935, South Africa \label{saao}
    \and Departamento de Física, Universidad Técnica Federico Santa María, Avenida España, 1680 Valparaíso, Chile \label{ufsm}
    \and Millennium Nucleus for Galaxies (MINGAL), Chile \label{mingal}
    \and South African Astronomical Observatory, 1 Observatory Road, Cape Town, 7925, Republic of South Africa. \label{saaos}
    \and Jodrell Bank Centre for Astrophysics, School of Physics and Astronomy,
    University of Manchester, Oxford Road, Manchester M13 9PL, UK. \label{jodrell}
    \and United Kingdom SKAO Regional Centre (UKSRC), Oxford Road, Manchester M13 9PL, UK. \label{ukskao}
    \and Finnish Centre for Astronomy with ESO, (FINCA), University of Turku, FI-20014 Turun yliopisto, Finland \label{finca}
    \and Tuorla Observatory, Department of Physics and Astronomy, University of Turku, FI-20014 Turun yliopisto, Finland \label{tobs}
    \and Turku Collegium of Science, Medicine and Technology (TCSMT), University of Turku, , FI-20014 Turun yliopisto, Finland \label{tcs}
    \and INAF-Astronomical Observatory of Capodimonte, Salita Moiariello 16, I-80131, Naples, Italy \label{INAF}
    \and Department of Astronomy, University of Wisconsin-Madison, 475 N Charter St, Madison, WI, 53706, USA \label{uwisc}
    \and Département de physique, de génie physique et d’optique, Université Laval, Québec (QC), G1V 0A6, Canada \label{laval}
    \and Western Sydney University, Locked Bag 1797, Penrith South DC, NSW 1797, Australia \label{wsu}
    \and Department of Physics and Astronomy, Macalester College, 1600 Grand Ave, St. Paul, MN 55105 USA \label{maca}
    }

   \date{Received ---; accepted ---}

 
  \abstract
  {
  We present new CO(1-0) measurements from ALMA and new \hi\ measurements from MeerKAT in four ram pressure stripped (RPS) galaxies in the core of the Hydra I cluster: NGC 3312, NGC 3314a, NGC 3314b and LEDA 753342. These deep data, with compatible spatial resolution in CO and \hi\ ($\sim15''$, $\sim$4 kpc), help increase the low statistics of galaxies observed in CO at extreme infall stage for probing how molecular and atomic gas behave under high ICM pressures. We studied the cold gas morphologies of the four Hydra I galaxies, and found asymmetries in their detected H$_2$ disks with respect to their \hi\ and optical disks, which could be related to the relatively high star formation rates of these galaxies. No traces of stripped CO were found in the tails, which could support the scenario that H$_2$ displays morphological and velocity asymmetries in the disks before being stripped. We then compared the H$_{2}$/\hi\ ratios and H$_{2}$ vs. \hi\ deficiencies between the four Hydra I galaxies, the xCOLDGASS sample dominated by field galaxies and VIVA-VERTICO Virgo RPS galaxies. LEDA 753342 ($\rm log~M_{\star}/M_{\odot} = 8.6$) is below the stellar mass limit of the reference samples ($\rm log~M_{\star}/M_{\odot} = 9$), but its $\rm M_{H_2}/M_{\hi}$ and cold gas deficiencies are compatible with those predicted by the lower stellar mass field and Virgo cluster galaxies. NGC 3312 and NGC 3314a are in an intermediate stage between field galaxies and more advanced RPS Virgo galaxies. NGC 3314b is more consistent with Virgo cluster galaxies more advanced in RPS, than with field galaxies. We reason that LEDA 753342 is at a very early stage of RPS, that NGC 3314a and NGC 3312 are at a similar timescale of early but very active RPS with differences potentially due to unequal initial total baryonic mass and infall angles, and that NGC 3314b is the latest stage in RPS in the sample. These galaxies display RPS across different evolutionary stages, and suggest that RPS starts affecting the CO reservoir before it starts being stripped from the galaxy.}


   \maketitle

\section{Introduction} \label{sec:Intro}

Ram pressure stripping \citep[RPS,][]{gun&gott1972} is a prominent hydrodynamical mechanism driving galaxy evolution in clusters \citep[see][]{cortese2021dawes,boselli2022}. The pressure exerted by the intracluster medium (ICM) on galaxies infalling within a cluster can remove their interstellar medium (ISM) in relatively short timescales \citep[$\sim$100–500 Myr;][]{abadi1999ram,mccarthy2008ram,2024GhoshRPS}, and act out to large cluster radii \citep[2–3R$_{200}$;][]{zinger2018quenching}. Both early simulations \citep[e.g.][]{tonnesen2009gas} and observations \citep[e.g.][]{chung2007virgo} agree that gas removal contributes to atomic gas deficiency and quenching of star formation in dense environments. 

RPS becomes more effective in massive cluster halos ($\rm M_h > 10^{14}~M_{\odot}$), where galaxies are generally more gas deficient compared to those in groups or the field \citep[e.g. ][]{brown2017cold}. RPS near the cluster core can reach pressures of $\rm 10^{-10}~g~cm^{-1}~s^{-2}$ \citep{roediger2005ram,2024GhoshRPS}. This is enough to strip up to $90\%$ of neutral gas for galaxies with masses in the range $ 9 < \rm log M_{\star}/M_{\odot} < 9.5$ \citep{xie2025impact}. Stripping efficiency will also depend on orbital parameters like inclination, infall direction, and pericentric passage, with radial orbits often leading to stronger gas loss \citep{chung2009vla,jaffe2015budhies,yoon2017history,boselli2022,salinas2024constraining}.

 The atomic gas of a galaxy is the most exposed to the ICM and the earliest tracer of RPS  \citep{chung2009vla,yoon2017history,chen2020ram,merluzzi2024ram}. The cause is that the acceleration of a gas cloud {\fontfamily{qag}\selectfont a} $\rm \ =P_{ram}/\Sigma$ is inversely proportional to the gas surface density $\Sigma$, which decreases in the outer galactic disk where the gas is additionally less gravitationally bound and consequently more vulnerable to RPS. Visual inspection has been the primary identification method of RPS candidates, traditionally for optical imaging \citep[e.g.][ and references therein] {poggianti2016jellyfish,salinas2024constraining}, but increasingly for \hi\ observations as well \citep[e.g.][]{yoon2017history,wang2021wallaby}. Nearby clusters such as Virgo, Antlia, Coma, A1367, Hydra I and Fornax have a \hi\ census of RPS galaxies \citep{chung2009vla,hess2015kat,Gavazzi2018b_rpsComa,Scott18_rpsA1367,molnar2022westerbork,Serra23_Fornax_HI_survey}, and agree that the extent of disc truncation correlates with \hi\ deficiency \citep{boselli2006environmental} and contributes to the quenching of star formation before RPS starts affecting \hi\ gas inside the optical radius \citep{cortese2021dawes,boselli2022}. 

In contrast, molecular gas is usually clumpier while its bulk is more centrally concentrated within the optical disk, where the gravitational restoring force is higher. These factors hinder the stripping of central molecular gas by the harsh environmental drag. Since the stripping timescale is longer than the lifetime of more external GMCs \citep[$\sim$5-50 Myr,][]{2012Miura,2025Ni_gmcs}, when stripped, molecular gas is removed as an entity alongside the \hi. Depending on the orientation of the ICM wind, the central molecular gas might be more easily affected by ram pressure in diverse ways, although this topic is still the object of ongoing research. The removal of \hi\ can expose the more central H$_2$ to the ICM \citep{2016Cortese_rps,cramer2020alma}, and hinder the conversion of H$_2$ by removing the atomic gas that otherwise could have collapsed into molecular gas. Alternatively, the pressure of the ICM can compress H$_2$ along the leading side of the galaxy, favoring its collapse into new stars \citep[e.g. ][]{cramer2020alma,roberts2022lotss}. Since molecular gas is tightly linked to star formation, the later can be used as a proxy to predict regions where dense molecular gas may be found. The surface density threshold $\sim1~$M$_{\odot}~$pc$^{-2}$ often defines the limit above which star formation has been observed in galaxy disks \citep[e.g.][]{2004Schaye_sflimit,2019MartinezLombilla_sflimit}. This threshold roughly corresponds to 1.26$\times10^{20}$cm$^{-2}$ for a Milky Way conversion factor $\rm \alpha_{CO}$ = 4.35 $\rm M_{\odot}pc^{-2}(Kkms^{-1})^{-1}$ \citep{2013Bolatto}, and can be considered a lower limit for the column densities at which CO is expected to be found. 

The detections of molecular gas in RPS galaxies agree in general with the scenario where star formation is affected by RPS even before it reaches central molecular gas, that is less affected by the ICM drag during early stages of ram pressure. Examples of resolved targeted CO surveys in clusters analyzing RPS include the resolved ALMA (Atacama Large Millimeter/submillimeter Array), the Fornax Cluster Survey \citep{zabel2019alma} and VERTICO \citep[Virgo Environment Traced in CO, ][]{brown2021vertico} survey in Virgo, or the unresolved Antlia CO (2$-$1) Survey \citep{cairns2019large} and CO Herschel Reference Survey \citep{boselli2014IIIcold} which covers the Virgo cluster as well as isolated objects. 
\cite{roberts2023vertico} observed that $\rm H_{2}$ asymmetries are smaller than the \hi\ asymmetries in a sample of star-forming satellite VIVA \citep[VLA Imaging of Virgo in Atomic gas, ][]{chung2009vla} and VERTICO \hi-tailed galaxies. They additionally suggested that CO-traced $\rm H_2$ over-densities trace star formation rate (SFR) excess in the leading half of the RPS galaxies. \cite{moretti2020high} suggest that ram pressure can increase the efficiency of \hi-to-H$_2$ conversion once it affects the discs of RPS galaxies. Simulations have also illustrated how RPS can remove diffuse molecular gas from the disk of galaxies, while compressing the leading half of the galaxy and increasing the conversion of \hi\ into $\rm H_2$ that boosts the SFR of the galaxy \citep{lee2020_simu}. This complements the results from \cite{brown2023vertico}, who showed that star formation quenching in \hi-poor Virgo satellite galaxies is caused by a simultaneous reduction in the molecular gas surface densities and boost of star-forming efficiency at fixed stellar mass surface density. All these effects combine to ultimately quench the galaxy. However, all these results offer a limited understanding of how RPS affect molecular gas that is largely based on studies of Virgo galaxies or optically selected galaxies. To build a more complete picture, it is necessary to investigate molecular gas across a broader range of galaxy populations and cluster environments.

\begin{figure}[t]
\includegraphics[width=\columnwidth]{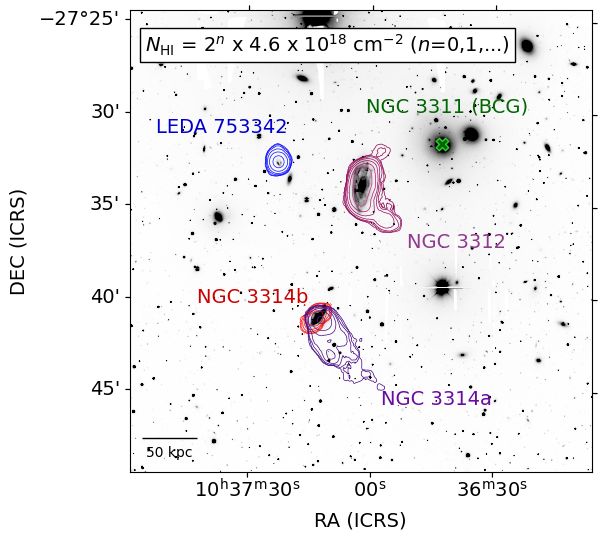}
\centering
 \caption{Grayscale DECam \textit{g}-band map of Hydra I core with a surface brightness limit of $\mu( 3\sigma;10''\times10'')$=26.9 mag $\rm arcsec^2$. MeerKAT \hi\ contours of the RPS galaxies NGC 3312, NGC 3314a, NGC3314b, and LEDA 753342 are colored. NGC 3311 is the brightest cluster galaxy.}
    \label{fig:map_core}
\end{figure}

The Hydra I cluster is a nearby cluster \citep[z $=$ 0.012, corresponding to a distance of d $=$ 58.6 Mpc;][]{Tully2015} with a recent history of galaxy accretion. In terms of mass \citep[3.02$\rm \times10^{14}~M_{\odot}$;][]{reiprich2002mass} and relaxation, Hydra I lies in an intermediate stage between the rich, unrelaxed Virgo cluster \citep[6.3$\pm$0.9$\times$10$^{14}$ M$_{\odot}$;][]{2020Kashibadze} and the not fully relaxed, relatively low mass \citep[$\rm \sim 4.5 \pm 1\times 10^{13}$ M$_{\odot}$, ][]{2001Drinkwater} Fornax cluster. In Virgo, RPS efficiently removes extended \hi\ \citep{chung2009vla}, leaving galaxies with sharply truncated atomic disks \citep{2004Kenney, 2009Vollmer} while their inner CO-traced molecular gas is less stripped \citep{2008Vollmer}, though recent surveys show even molecular gas becomes truncated in the most stripped systems \citep{brown2021vertico,Zabel2022}. The less massive Fornax cluster also shows environmental suppression of gas, manifesting as low CO masses and disturbed or reduced molecular reservoirs \citep{zabel2019alma,2021Loni}, consistent with a cluster still actively assembling and with active \hi\ stripping \citep{Serra23_Fornax_HI_survey}. In contrast, Hydra I has a more regular X-ray morphology and intermediate mass \citep{Fitchett1988}, with galaxies that display moderate gas deficiencies at best \citep{McMahon92,denes2014_hidef}. This suggests a cluster whose core is relatively settled and defined as a weak cool core clusters \citep{Srivastava2025_Xhydra}, and whose environment has not yet removed gas as efficiently as Virgo. Thus, Hydra I represents a transitional environment less extreme than the harsh stripping regime in Virgo, but more evolved and gas-rich than the lower-mass Fornax cluster \citep{2004Bohringer,2008Misgeld}.

Studies focused in \hi\ such as \cite{wang2021wallaby} have observed that at least 70\% of all the members of Hydra I might be undergoing RPS. \cite{delaCasa2025} combined optical and \hi\ data and observed that infalling galaxies experience a burst in star formation potentially related to RPS. Using deep \hi\ and optical observations, \cite{Hess2022} analyzed central Hydra I galaxies NGC 3312, NGC 3314a/b and LEDA 753342 and confirmed they are undergoing active RPS. They suggested that the relative impact of RPS on each galaxy would depend on factors such as belonging to a substructure that could decrease the impact of the ICM.  Recent optical studies in Hydra I such as \cite{lima2021environmental}, \cite{LaMarca2022} and \cite{Spavone2024_HydraStructure} have found traces of at least three substructures in the core (encompassing the brightest cluster galaxy NGC 3311), to the N and to the SE of the former substructure. \cite{Spavone2024_HydraStructure} reported that the SE region of Hydra I hosts the highest number of RPS galaxies. \cite{delaCasa2025} observed that the three potential substructures in Hydra I are aligned with the filaments connecting it to other clusters, and may be related to the distribution of galaxies with enhanced SFR and undergoing RPS, particularly in the SE region.    

 We present the first CO view of galaxies near the core of the Hydra I cluster, combined with new deep \hi\ observations. We use targeted ALMA CO(1-0) observations and a deep MeerKAT 32K \hi\ mosaic of four RPS galaxies near the core of the Hydra I cluster, namely: NGC 3312, NGC 3314a/b and LEDA 753342. These galaxies belong to the SE young foreground substructure \citep{Fitchett1988,MacMahon1993, Hess2022,Spavone2024_HydraStructure} undergoing strong RPS due to its proximity to the cluster core (Fig. \ref{fig:map_core}). The galaxies cover a wide stellar mass range of $\rm 8.5<log(M_{\star}/M_{\odot})<11.2$ and have relatively high star-forming rates (see Table \ref{table:logSmass}) that could be related to ongoing RPS as they fall into the cluster \citep{delaCasa2025}. In this work we link neutral and molecular gas morphologies, compare the cold gas reservoirs and their deficiencies with field galaxies and galaxies from other clusters, and evaluate these results to explore quenching pathways and their observed SFRs. 

 We consider a $\rm \Lambda CMD$ cosmology with $\Omega_{\rm m}= 0.27$, $\Omega_{\Lambda} = 0.73$, and H$_0 = 70$ km s$^{-1}$ Mpc$^{-1}$. We assume a redshift of $z_{\rm HC}\approx0.012$ for the Hydra I cluster, corresponding to a distance of d$_{\rm HC}=58.6$ Mpc \citep{Tully2015} and a characteristic radius of R$_{200}=1.35$ Mpc containing a mass of M$_{\rm 200}=3.02\times10^{14}$~M$_{\odot}$ \citep{reiprich2002mass}. We assume that the cluster core is at the location of NGC 3311, the brightest cluster galaxy: $\rm RA_{core}=159.17^{\circ}$, $\rm DEC_{core}=-27.52^{\circ}$. The systemic velocity adopted for Hydra is $\rm v_{sys}=3825km/s$ and its velocity dispersion is $\rm \sigma_{HC}=620km/s$ \citep{wang2021wallaby}.
 
 The outline of this work is as follows. We outline the data in Sec. \ref{sec:Data} and describe the gas distribution and morphologies for every galaxy in Sec. \ref{sec:results}. We compare cold gas fractions and deficiencies in Sec. \ref{sec:gasfracs}, placing them in the context of other clusters. We discuss these findings in the context of RPS in Sec. \ref{sec:discussion}, and summarize key results in Sec. \ref{sec:summary}.

\section{Data}\label{sec:Data}

\subsection{CO observations}
We present our observations in ALMA Band 3 \citep{claude2008performance} of the CO(1-0) line at 115.27 GHz in the Hydra I cluster (project code 2022.1.00716.S, PI K.~Hess) for NGC 3312, NGC 3314a/b and LEDA 753342. These observations were performed with the ALMA Compact Array (ACA) as part of ALMA Cycle 9, and provide a spatial resolution of 11$^{\prime\prime}$ (3.1 kpc) for the CO(1-0) line. The choice of the CO(1-0) line is motivated by lower uncertainties associated to the estimation of H$_2$ relative to higher order transitions, as well as the excellent match between the spatial resolution of ACA for 115 GHz and that of the observations from MeerKAT at 1.4 GHz of 16$^{\prime\prime}$x11$^{\prime\prime}$.
 
 Of these four galaxies, LEDA 753342 and NGC 3314b were covered with one pointing, while NGC 3312 and NGC3314a required two and three pointings, respectively. Multiple pointings of the same galaxy were mosaicked to cover the disk and the RPS star-forming tails identified both in \hi\ and H$\alpha$.
 
We used the ALMA Science Pipeline Software \citep{hunter2023alma} to reduce the data and then used the Common Astronomy Software Applications \citep[CASA;][]{mcmullin2007casa} for further imaging. We used the {\tt tclean} task to re-process the maps with \textit{natural weighting}, \textit{Högbom deconvolver}, and a \textit{width} of 10 channels. To promote detections, the final cubes have a generous spectral resolution of 65\kms and were smoothed from the original 14$^{\prime\prime}$x11$^{\prime\prime}$ beamsize to 25$^{\prime\prime}$x25$^{\prime\prime}$. To favor later source finding even more we also lowered the \textit{noise threshold} {\tt tclean} parameter to 4.25, the \textit{minbeamfrac} parameter to 0.1, and the \textit{sidelobe threshold} and \textit{low noise threshold} parameters to 1.0. The average RMS for the resulting cubes is 0.015 Jy/beam.

We applied the Source Finding Application \citep[SoFiA-2,][]{westmeier2021sofia} in each cube/mosaic to find sources. False detections with integrated signal-to-noise ratio (S/N) below 5 were removed, leaving true detections for galaxies NGC 3312 and NGC 3314a/b and no reliably detected CO for LEDA 753342.

\begin{center}
    \begin{table*}[t]
    \centering
    \caption{\label{table:logSmass}Parameters of the four Hydra I galaxies studied in this work}
    \begin{tabular}{ c c c c c c c c c c} 
    \hline
       Galaxy & $\rm log(M_{\star})$  &   $\rm SFR_{\rm H\alpha}$ & $\Delta$MS & $\rm \alpha_{CO}$ &$\rm log(M_{\rm H_{2}})$ & Def($\rm H_2$) &$\rm log(M_{\rm H_{I}})$ & Def($\rm H_I$)\\

         & $\rm log(\rm [M_{ \odot}])$ &   $[\rm M_{\odot}/yr]$ &  &   \underline{[$\rm M_{\odot}pc^{-2}$]} &$\rm log([\rm \rm M_{ \odot}])$ & [dex] & $\rm log([\rm M_{ \odot}])$ & [dex] \\

          &  &   &  &   [$\rm Kkms^{-1}$] & &  &  &  \\

     \hline
       NGC 3312 & 11.12$\pm$0.08 \textcolor{cyan}{[1]}& 12.45$\pm$2.40 & 0.48 & 3.55 & 9.44$\pm$0.04 & 0.06$\pm$0.30 & 9.69$\pm$0.01 & 0.87$\pm$0.22\\
       NGC 3314a & 9.9$\pm$0.1 \textcolor{cyan}{[2]}& 1.01$\pm$0.11 & 0.20 & 4.14 & 8.86$\pm$0.12 & -0.09$\pm$0.32 & 9.27$\pm$0.01 & 1.31$\pm$0.23\\
       NGC 3314b & 10.5$\pm$0.1 \textcolor{cyan}{[2]}& 3.95$\pm$0.11 & 0.34 & 2.76 & 9.42$\pm$0.07 & -0.28$\pm$0.31 & 8.69$\pm$0.01 & 1.89$\pm$0.27\\
       LEDA 753342 & 8.6$\pm$0.1 \textcolor{cyan}{[1]}& 0.11$\pm$0.02 & 0.32 & 9.57 & $<$7.7$\pm$0.5 & $>$0.28$\pm$0.60 & 8.77$\pm$0.01 & 0.23$\pm$0.24\\
     \hline
    \end{tabular}\\
    {Column 1: Name of the galaxy. Column 2: Stellar mass from \textcolor{cyan}{[1]} \cite{delaCasa2025} and \textcolor{cyan}{[2]} \cite{Hess2022}, assuming the flux ratio of the NGC 3314a/b is 80\%/20\% of total flux. Column 3: Star formation rate from \cite{delaCasa2025}. Column 4: Distance to Main Sequence from \cite{delaCasa2025}, estimated as $\rm \Delta MS = SFR_{H\alpha}-SFR_{MS}$ where $\rm SFR_{MS}$ is the star formation for a galaxy of the same M$_{\star}$ on the main sequence. Column 5: CO-to-H$_2$ conversion factor derived following \cite{accurso2017deriving}. Column 6: Molecular gas mass. Uncertainties do not take into account the uncertainties on the conversion factors. Column 7: Molecular gas deficiency. Column 8: \hi\ mass. Column 9: \hi\ deficiency. Columns 6-9 contain estimations performed for this work following the procedure described in App. \ref{appendix:h2mass},\ref{appendix:himass}}.
    \end{table*}
\end{center}

\subsection{\hi\ observations}

We make use of a new set of measurements of the 1.4 GHz \hi\ spectral line with MeerKAT in the 32K mode (project ID SCI-20210212-NM-01). These observations are based on a single 10 hr pointing of the center of the Hydra I cluster in the L-band, with a resolution of 26.123 kHz (5.5 km/s at z=0).  They achieve a spatial resolution of 11"x16" (3.1x4.1 kpc), with a \textit{Briggs robust weighting} of 0.5. This resolution is well matched to the ALMA ACA resolution (11"; 3.1 kpc), allowing to compare the molecular and neutral gas morphologies over similar spatial scales.

The data reduction processes followed a similar strategy to \cite{serra2019} and \cite{healy2021} and will be detailed in Makda et al. (in prep.). The data reduction was performed on the ilifu cloud computing facility, using the {\tt CARACal} pipeline \citep{jozsa2020caracal,jozsa2020meerkathi} to calibrate visibilities. The {\tt TRICOLOUR} flagger and {\tt AOFLAGGER} packages were used to flag RFI in the calibrator and target field respectively. The cross-calibrated visibilities of the Hydra I field were imaged with {\tt WSClean} using a mask created by {\tt breizorro}, and self-calibrated with {\tt CubiCal}. We used Multi-frequency synthesis in {\tt WSClean} to image the continuum and perform continuum subtraction. {\tt WSClean} was then used to create the \hi\ cube with a \textit{robust weighting} of 0.5 and \textit{pixel scale} of 3$^{\prime\prime}$. The cleaning strategy of detected \hi\ sources followed \cite{healy2021}, implementing SoFiA-2 to create masked \hi\ sources and {\tt WSClean} to clean the sources. The high resolution cubes used in this work have an average rms noise of 0.35 mJy beam$^{-1}$ channel$^{-1}$ which translates to a 1$\sigma$ \hi\ column density of $\rm N_{HI} = 1.2 \times 10^{19}~cm^{-2}~channel^{-1}$. The cleaned \hi\ cubes were smoothed to 22$^{\prime\prime}$ and 45$^{\prime\prime}$ to detect the most diffuse gas. They were then primary beam corrected, followed by source detection and characterization with SoFiA-2.

\subsection{DECam optical observations}\label{sec:DECam_data}
The ancillary optical data employed in this study comes from the 
Dark Energy Camera installed on the CTIO Blanco 4m telescope \citep[DECam; ][]{flaugher2015dark}, as part of project 2021A-0117 (PI: R. Kotulla). The data consist of optical imaging of the Hydra I cluster with a surface brightness limit of $\mu( 3\sigma;10^{\prime\prime}\times10^{\prime\prime})$=26.9 mag $\rm arcsec^2$ in the \textit{g} band. The observations were first presented in \cite{Hess2022} and further described and analyzed in \cite{delaCasa2025}. We refer to these past works for a complete description of the data treatment including data reduction, normalization and scaling, and continuum subtraction.

For this work, we combine \textit{r}, \textit{g} and \textit{u} bands to build the color composites of the galaxies in question. H$\alpha$ images are also used to visualize the regions with active star formation. SFRs and distances to the Main Sequence ($\Delta$MS) in this work are taken from \cite{delaCasa2025} and can be found in Table \ref{table:logSmass}. They were estimated converting optical H$\alpha$ luminosities using Eq. (2) from \cite{Kennicut1998_sfrha}, and applying a dust correction based on stellar mass derived in \cite{garn2010dustcorrection}. 

\subsection{Control sample}
Through this work we use the xCOLDGASS catalogue \citep[Extended CO Legacy Database for GASS, ][]{saintonge2017xcold} as a reference for the cold gas content of field galaxies. This amounts to 305 galaxies with \hi\ observations from Arecibo (as part of the xCOLDGASS program) or from the ALFALFA survey \citep[Arecibo Legacy Fast ALFA survey, ][]{Haynes2011} when the latter was not available, and CO observations from the IRAM (Institut de Radioastronomie Millimétrique) 30m telescope. To provide a reference for resolved gas in cluster galaxies, we additionally use Virgo cluster galaxies from the VIVA \citep{chung2009vla} and VERTICO \citep{brown2021vertico} surveys as comparison. We have selected 43 galaxies that are present in the \hi\ VIVA and CO-traced H$_2$ VERTICO catalogues.

\subsection{Ancillary data}

We make use of a MUSE (Multi-Unit Spectroscopic Explorer) cube from the LEWIS program \citep[Looking into the faintEst WIth MUSE, ][]{Iodice2023_LEWIS,hartke2025looking} in section \ref{sec:3314a_results} to compare the velocity of the H$\alpha$ emission in the filament of NGC 3314a to the cospatial neutral gas. We also make use of publicly available Hubble Space Telescope (HST) images \citep{keel2001seeing} in section \ref{sec:discussion}.

\section{Results}\label{sec:results}

\subsection{Cold gas distribution and morphology}  

\subsubsection{NGC 3312}

\begin{figure*}[t]
    \centering
    \begin{minipage}[c]{0.79\textwidth}
        \centering
    \includegraphics[width=\textwidth]{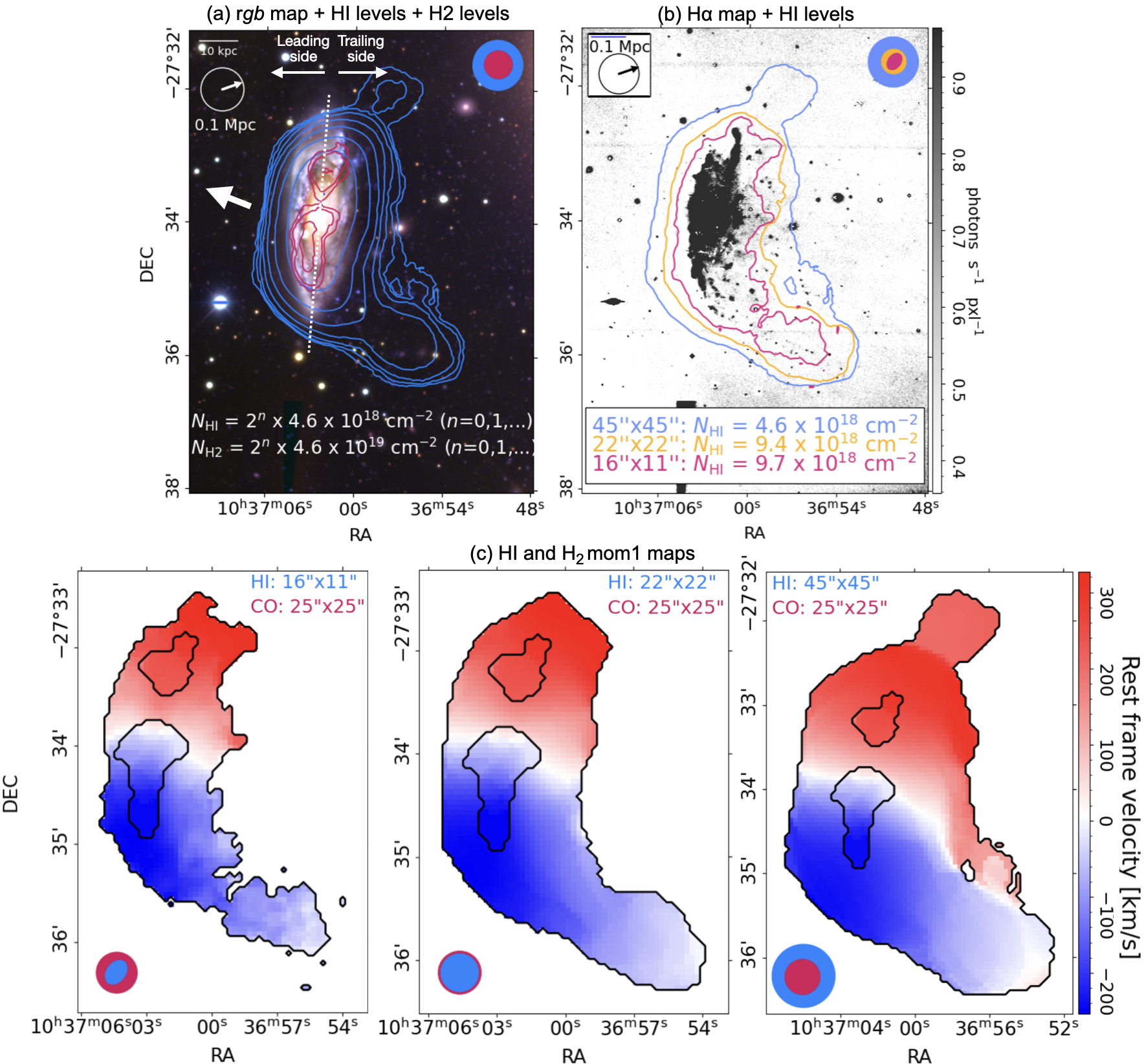}
    \end{minipage}
    \hfill
    \begin{minipage}[c]{0.2\textwidth}
        \caption{ Spiral galaxy NGC 3312. (a) Color composite of the \textit{r}, \textit{g}, and \textit{u} bands from DECam. In red, $\rm H_2$ ALMA CO(1-0) contours. MeerKAT \hi\ contours are colored blue and delimit the extent of the neutral gas in the ram pressure stripped tail of NGC 3312. The CO and \hi\ beams have been smoothed to 25$^{\prime\prime}$x25$^{\prime\prime}$ and 45$^{\prime\prime}$x45$^{\prime\prime}$, respectively (top right corner). In the top left of the figure there is a scale of 10 kpc and a compass indicating the direction to the brightest cluster galaxy NGC 3311, that we assume is at the center of Hydra I. The distance below the compass is the projected distance to the cluster core. The thick white arrow indicates the suggested direction of movement. (b) Greyscale DECam $\rm H\alpha$ image covering NGC 3312 with contours representing the 3$\sigma$ \hi\ levels of NGC 3312 at different spatial resolutions. The size of the beam at every resolution is displayed in the top right corner. (c) Moment 1 velocity maps of detected CO, and the different \hi\ resolutions in panel (b). \label{fig:ngc3312}}
    \end{minipage}
\end{figure*}

Figure \ref{fig:ngc3312} shows the morphology and kinematics of NGC 3312 ($\rm v_{rec,HI} = 2818$km/s) in different wavelengths. Fig \ref{fig:ngc3312} \textcolor{cyan}{a} displays a DECam \textit{rgb} composite made with DECam \textit{r}, \textit{g} and \textit{u} bands. Blue \hi\ contours have angular resolution of 45$^{\prime\prime}$x45$^{\prime\prime}$, or $\sim 8.64$ kpc at a redshift of $\rm z =0.009$ \citep{delaCasa2025}, and minimum column densities of $\rm N_{HI}\sim5\times10^{18} ~cm^{-2}$. Red CO(1-0) contours have angular resolution of 25$^{\prime\prime}$x25$^{\prime\prime}$, or $\sim 4.8$ kpc at a redshift of $\rm z =0.009$, and column densities as low as $\rm N_{H_2}\simeq10^{20}~cm^{-2}$. 

The \hi\ gas content reveals two tails exceeding the stellar disk to the west of the galaxy. We assume that the direction of motion is opposite to the tails (thick arrow in Fig. \ref{fig:ngc3312} \textcolor{cyan}{a}). The first tail follows an unwinding spiral arm in the northern half of the galaxy. This arm is blue on the side most exposed to the wind and dust obscured on the downwind side, suggesting that dust may be being stripped as well. Figure \ref{fig:ngc3312} \textcolor{cyan}{b} shows the $\rm H\alpha$ images of this galaxy, evidencing regions of active and unobscured star formation in the upwind side of this arm and traces of dust in the downwind side.

Fig. \ref{fig:ngc3312} \textcolor{cyan}{b} also displays the 3$\sigma$ \hi\ contours at different resolutions over the H$\alpha$ image from DECam. Although the \hi\ shape traces the star-forming regions including the upper spiral arm, the northern tail is only revealed at the lowest spatial resolution. The second tail extends to 40 kpc at column densities of N$_{\hi}\simeq10^{19}$ cm$^{-2}$, covering the full length over which we find H$\alpha$ emission tracing star formation.

Fig. \ref{fig:ngc3312} \textcolor{cyan}{c} displays the moment 1 velocity maps of the three resolutions shown in Fig. \ref{fig:ngc3312} \textcolor{cyan}{b}. High resolution \hi\ is to the left, with increasing smoothing to the right. The dimmest \hi\ gas is only visible in the rightmost panel, where the \hi\ beam is of 45$^{\prime\prime}$x45$^{\prime\prime}$, showing how the two tails follow the rotation of the disk. The central molecular gas also follows the rotation of the atomic gas in the disk. The left panel in Fig. \ref{fig:ngc3312} \textcolor{cyan}{c} shows how the molecular gas distribution in NGC 3312 is not centered in the \hi\ disk. It is displaced towards the leading spiral arm and the northern unwinding arm, where \hi\ contours are roughly above $\rm N_{HI}=6.4\times10^{20}~cm^{-2}$. This occurs potentially because the increased pressure and dust presence promote \hi\ to H$_{2}$ conversion \citep{moretti2020gasp}. If we divide the galaxy in two halves through its major axis (see App. \ref{App:ngc_3312} for more details), we find that 47\% of the total CO traced in the galaxy is in the leading half and the other 53\% in the trailing half. However, 33\% of the detected CO is in the unwinding arm, leaving the majority of the remaining molecular gas in the leading edge of the galaxy (40\%). The rest of the CO (27\%) is found in the trailing half, to the south of the unwinding arm, where the tail is originated and the H$\alpha$ emission tracing recent star formation is high as well.

\begin{figure*}[t]
\includegraphics[width=\textwidth]{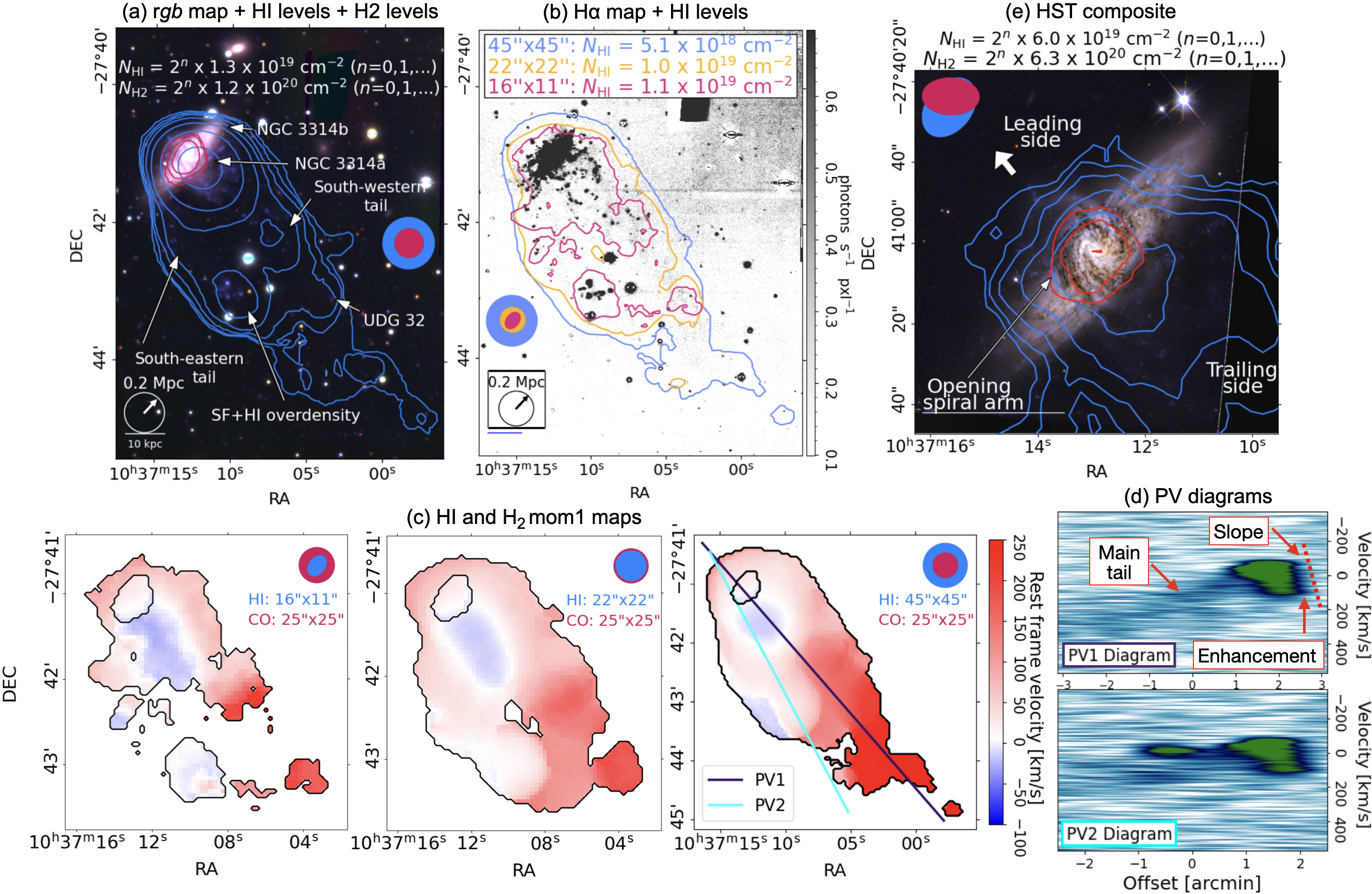}
\centering
 \caption{ Face-on spiral galaxy NGC 3314a (foreground), in chance projection with spiral galaxy NGC 3314b (background). (a) \textit{Rgb} color composite, same specifications as Fig. \ref{fig:ngc3312}. (b) Greyscale DECam $\rm H\alpha$ image covering NGC 3314a (foreground) and NGC 3314b (background) with contours representing the 3$\sigma$ \hi\ levels of NGC 3314a at different resolutions. (c) Moment 1 velocity maps of detected CO, and \hi\ with increasing smoothing from left to right, with the beamsize represented in the top right corner. (d) \hi\ PV diagrams in the direction of the colored lines in the moment map. (e) HST \textit{rgb} composite made with the F450W, F675W, and F814W filters \citep{keel2001seeing}. Overlaid we have represented the highest resolution \hi\ ($16^{\prime\prime}\times11^{\prime\prime}$, blue) and CO(1-0) ($14^{\prime\prime}\times9^{\prime\prime}$, red) contours. The white arrow indicates the suggested direction of movement.}
    \label{fig:ngc3314a}
\end{figure*}

\subsubsection{NGC 3314a} \label{sec:3314a_results}

Figure \ref{fig:ngc3314a} displays the face-on foreground galaxy NGC 3314a ($\rm  v_{rec,HI}=2817$\kms) orbiting to the northwest, and in chance projection with the inclined background galaxy NGC 3314b ($\rm  v_{rec,HI}=4657$\kms) that moves towards the southeast, as was discussed in \cite{keel2001seeing} and \cite{Iodice2021_ngc3314}. As these works, in this study we assume that there are no tidal interaction between the aligned galaxies given the lack of tidal features in the stellar body. Fig. \ref{fig:ngc3314a} \textcolor{cyan}{a} shows the \textit{rgb} color composite with the \hi\ and H$_{2}$ contours overlaid in blue and red, respectively. The \hi\ tail hosts two distinct filaments of star formation that can also be seen in the H$\alpha$ image from Fig. \ref{fig:ngc3314a} \textcolor{cyan}{b}. The highest \hi\ spatial resolution in Fig. \ref{fig:ngc3314a} \textcolor{cyan}{b} traces these filaments as well. Our deep observations reveal the tail out to $\sim 50$ kpc in projection for the 25$^{\prime\prime}$x25$^{\prime\prime}$ smoothed data, and $\sim 70$ kpc for the 45$^{\prime\prime}$x45$^{\prime\prime}$ smoothed data. 

Fig. \ref{fig:ngc3314a} \textcolor{cyan}{c} shows the CO-traced H$_2$ and \hi\ moment 1 velocity maps with the different \hi\ spatial resolutions in Fig. \ref{fig:ngc3314a} \textcolor{cyan}{b}. The molecular gas is only found in the central disk of the galaxy, and follows the rotational pattern of the \hi. The northern more redshifted edge of the \hi\ could be linked to gas removal that gives rise to the southwestern tail. This is more evident in the PV diagrams from Fig. \ref{fig:ngc3314a} \textcolor{cyan}{d}, performed along the two cuts in the directions indicated in the right panel of Fig. \ref{fig:ngc3314a} \textcolor{cyan}{c}. The first one traces the southwestern tail, while the second one traces the shorter southeastern tail. The removal follows an enhancement in the PV diagram at rest frame velocity $\sim100$km/s, which extends to higher velocities in PV1 below the main tail. The moment 1 velocity maps also suggests that the southeastern, less redshifted HI tail overlaps with a spiral arm and is potentially following the removal of gas from said arm. In PV2 (Fig. \ref{fig:ngc3314a} \textcolor{cyan}{d}) we see the detachment of this material at rest frame velocities $\sim0$km/s.

Figure \ref{fig:ngc3314a} \textcolor{cyan}{e} shows a color composite made with HST imaging from \cite{keel2001seeing}. The distribution of dust in NGC 3314a is highlighted by the background emission from NGC 3314b. Contours represent the highest resolution atomic and molecular gas levels. CO is predominantly found where the \hi\ is above $\rm N_{\hi}\sim1.76\times 10^{20}~$cm$^{-2}$, tracing the dustier regions in the spiral arms. The lowest contour is slightly extended towards the south-eastern direction in which the leading spiral arm is opening and creating the southeastern \hi\ tail. Molecular gas too diffuse to be traced by the current CO observations could be being dragged out of the disk along with the neutral gas. 

\begin{figure}[t]
\includegraphics[width=\columnwidth]{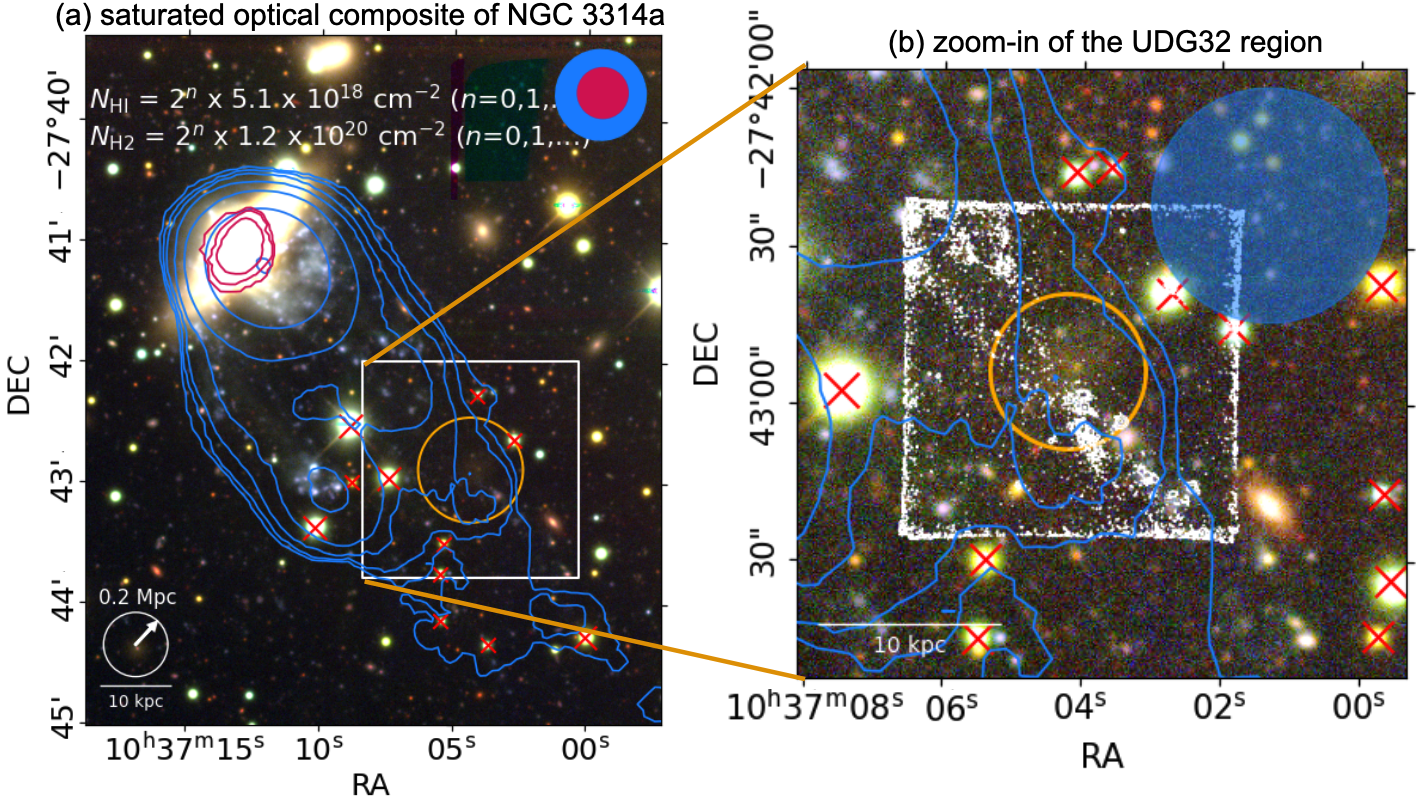}
\centering
 \caption{(a) Optical \textit{rgb} image with CO and \hi\ contours of the NGC 3314a/b alignment with saturated \textit{g} and \textit{u} bands. 'X' have been placed over the known stars. It evidences the two main RPS tails of NGC 3314a that follow the rotation of its arms. UDG 32 is in the center of the squared region, delimited by the orange circle. (b) Zoom-in of the squared region in Fig. \ref{fig:ngc3314a}, showing where the NGC 3314a gas tail overlaps with UDG 32. UDG 32 is inside the orange circle. The white contours show the $\rm H\alpha$ filament reported by \cite{hartke2025looking}, which is aligned in direction with the \hi\ contours.} \label{fig:ngc3314a_combo2}
\end{figure}

Even though no CO is detected in the tails, there is a SF+\hi\ overdensity in the southeastern one, located 20 kpc in projection south from the stellar disk with the same recessional velocity. The highest resolution velocity map (left panel of Fig. \ref{fig:ngc3314a} \textcolor{cyan}{c}) suggests that this clump might have rotation. The \hi\ column densities in the clump from the southeastern tail are of $\rm N_{HI}=3.2\times10^{20}$ cm$^{-2}$. We estimate an upper limit of gas content in the SF+\hi\ overdensity of $\rm M_{HI} = 10^{8}$ M$_{\odot}$ and $\rm M_{H_2} = 10^{7.07}$ M$_{\odot}$ (see App. \ref{app:nondets}). To estimate the M$_{H_2}$ we assumed the same $\rm \alpha_{CO}$ as the disk, which would be much lower than the one expected for such a low metallicity region in a tail of gas \citep[see App. C from][for an analysis of the abundance and metallicity of star-forming clumps inside MUSE FoV]{hartke2025looking}. The optical extent of the clump is $\sim10$ kpc, while the \hi\ envelope with column densities above $\sim 10^{20}$ cm$^{-2}$ (fourth contour in Figs. \ref{fig:ngc3314a} \textcolor{cyan}{a} and \ref{fig:ngc3314a_combo2} \textcolor{cyan}{a}) has a size of $\sim15$ kpc accounting for the beamsize. This size is similar to the diameter of the dwarf galaxies in Hydra included in the sample from \cite{delaCasa2025}.

The southwestern, more redshifted filament of the tail of NGC 3314a overlaps with an ultra-diffuse galaxy (UDG) named UDG32 \citep{iodice2021formation}. The origin of UDG32 is uncertain but RPS is one of the proposed scenarios to explain it \citep{hartke2025looking}. In Fig. \ref{fig:ngc3314a_combo2} \textcolor{cyan}{a} we show a saturated composite of the \textit{r}, \textit{g}, and \textit{u} DECam bands of NGC 3314a and its tail, where UDG32 is visible. In Fig. \ref{fig:ngc3314a_combo2} \textcolor{cyan}{b} we can appreciate how the stellar component of UDG32 (inside orange circle) and the \hi\ stripped material of NGC 3314a overlap with an H$\alpha$ filament observed with MUSE by \cite{hartke2025looking}, within the white square. We estimate a velocity of $v_{tail}=3033\pm67 \rm km/s$ for the co-spatial \hi\ overdensity, which is very similar to the velocity of the H$\alpha$ filament \citep[$v_{UDG32}=3085 \pm 120 \rm km/s$,][]{hartke2025looking}.

\subsubsection{NGC 3314b}

\begin{figure*}[t]
    \centering
    \begin{minipage}[c]{0.79\textwidth}
        \centering
    \includegraphics[width=\textwidth]{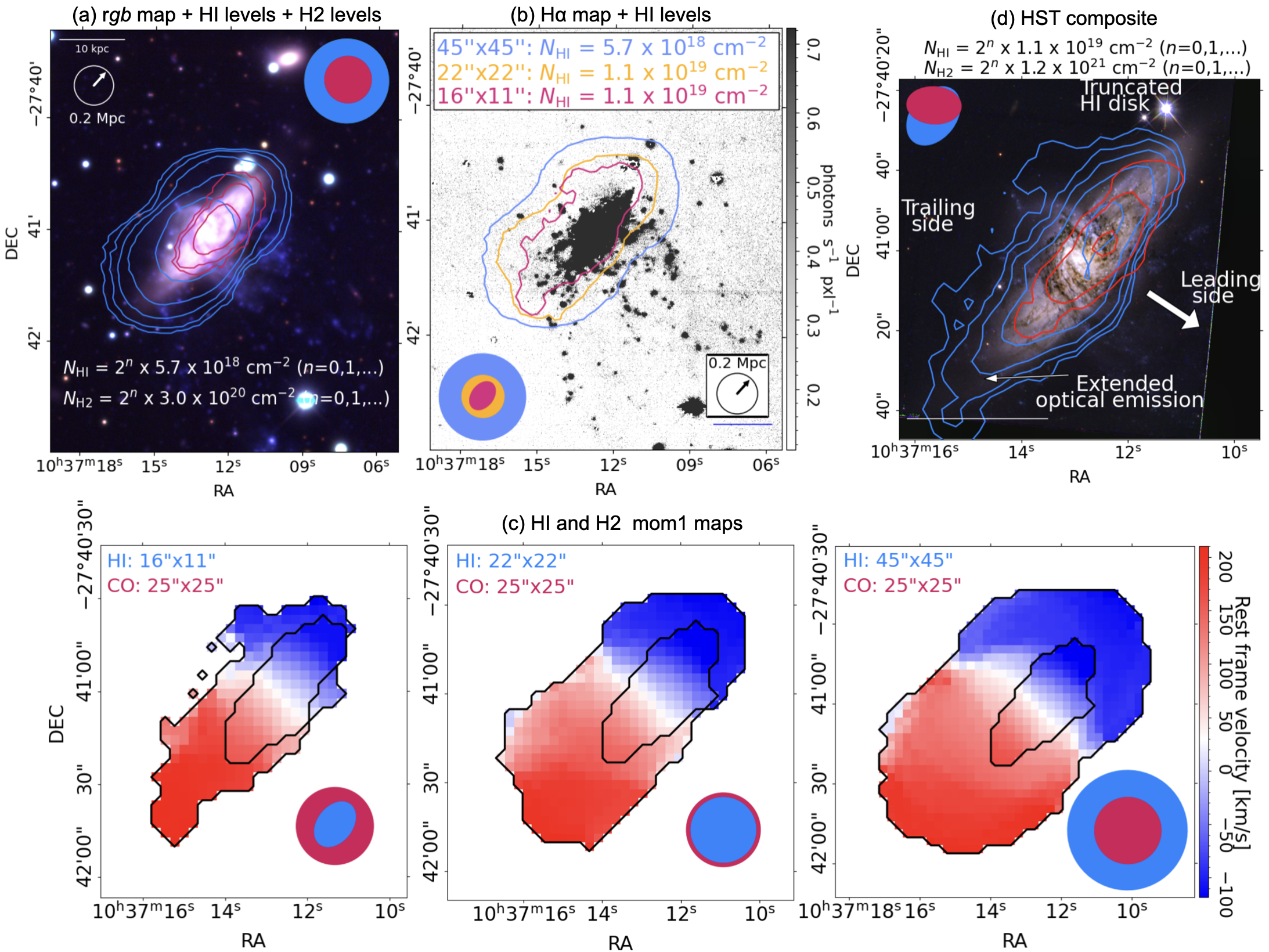}
    \end{minipage}
    \hfill
    \begin{minipage}[c]{0.2\textwidth}
        \caption{ Edge-on spiral galaxy NGC 3314b (background), in chance alignment with spiral galaxy NGC 3314a (foreground). (a) \textit{rgb} color composite, same specifications as Fig. \ref{fig:ngc3312}. (b) Greyscale DECam $\rm H\alpha$ image covering NGC 3314b (background) and NGC 3314a (foreground) with contours representing the 3$\sigma$ \hi\ levels of NGC 3314b at different resolutions. (c) Moment 1 velocity maps of detected CO, and \hi\ with increasing smoothing from left to right, with the beamsize represented in the bottom right corner. (d) HST \textit{rgb} composite made with the F450W, F675W, and F814W filters \citep{keel2001seeing}. Overlaid we have represented the highest resolution \hi\ ($16^{\prime\prime}\times11^{\prime\prime}$, blue) and CO(1-0) ($14^{\prime\prime}\times9^{\prime\prime}$, red) contours for NGC 3314b. The thick white arrow indicates the suggested direction of movement. \label{fig:ngc3314b}}
    \end{minipage}
\end{figure*}

In Fig. \ref{fig:ngc3314b}, we present images of the almost edge-on background galaxy NGC 3314b ($\rm v_{rec,HI}=4591$km/s). The smoothed \hi\ observations and velocity maps in Figs. \ref{fig:ngc3314b} \textcolor{cyan}{a}, \textcolor{cyan}{b}, \textcolor{cyan}{c} do not display asymmetries, which are only revealed by the highest spatial resolution in Figs. \ref{fig:ngc3314b} \textcolor{cyan}{b}, \textcolor{cyan}{d}. In the left velocity map of Fig. \ref{fig:ngc3314b} \textcolor{cyan}{c} and in Fig. \ref{fig:ngc3314b} \textcolor{cyan}{d} we observe that the \hi\ disk in this galaxy is extended to the south east, whereas the H$_{2}$ disk is extended to the north west with respect to the optical disk. The optical emission from the stellar disk also seems fainter and extended towards the south east, in contrast to the sharper north west edge. The H$\alpha$ image of Fig. \ref{fig:ngc3314b} \textcolor{cyan}{b} is too contaminated by the foreground NGC 3314a to see if the \hi\ contours of the remaining tail have associated star formation. The H$_2$ distribution suggests that molecular gas content is potentially enhanced on the leading (eastern) side of the galaxy, and perhaps being removed from the trailing (western) edge.

\subsubsection{LEDA 753342}

\begin{figure*}[t]
    \centering
    \begin{minipage}[c]{0.79\textwidth}
        \centering
    \includegraphics[width=\textwidth]{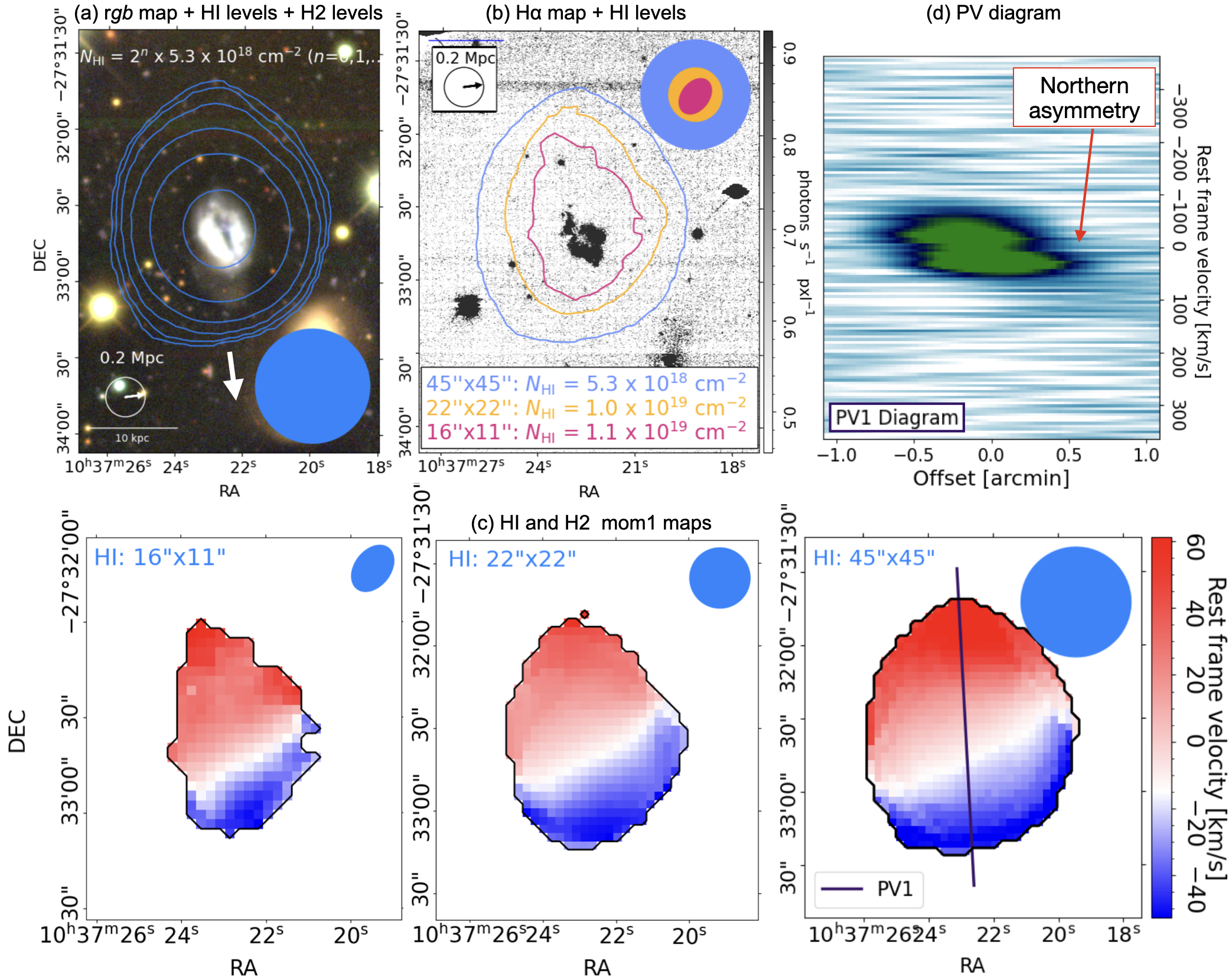}
    \end{minipage}
    \hfill
    \begin{minipage}[c]{0.2\textwidth}
        \caption{ Ring galaxy LEDA 753342 (a) Optical \textit{rgb} image with \hi\ contours of LEDA 753342 with a saturated \textit{g} band, evidencing residual light in the Southern half. No CO emission could be found in this case, so only MeerKAT \hi\ contours have been represented. The thick white arrow indicates the suggested direction of movement. (b) Greyscale DECam $\rm H\alpha$ image covering LEDA 753342 with contours representing the 3$\sigma$ \hi\ levels at different resolutions. (c) Moment 1 velocity maps of detected \hi\ for the different resolutions in panel (b). The line represents the direction of the PV slice. (d) \hi\ PV diagram in the direction of the colored line in the moment 1 velocity map. \label{fig:leda753342}}
    \end{minipage}
\end{figure*}

Figs. \ref{fig:leda753342} \textcolor{cyan}{a}, \textcolor{cyan}{b} show the optical and \hi\ profiles of LEDA 753342 ($\rm v_{rec,HI}=2716$ km/s). Fig. \ref{fig:leda753342} \textcolor{cyan}{c} displays the CO and different \hi\ resolution velocity maps of this galaxy. No CO detection was found for this galaxy above $\rm N_{H_2} \sim 10^{20} ~cm^{-2}$. In Fig. \ref{fig:leda753342} \textcolor{cyan}{a} we observe residual light to the south LEDA 753342, that could indicate a past interaction or collision (see Section \ref{sect:insights}). This emission is opposite to a slight \hi\ asymmetry in all spatial resolutions displayed in Figs. \ref{fig:leda753342} \textcolor{cyan}{b} and \ref{fig:leda753342} \textcolor{cyan}{c}, which suggests that the galaxy is moving towards the south. Fig. \ref{fig:leda753342} \textcolor{cyan}{d} is a \hi\ PV diagram across the longest axis of the \hi\ disk, showing emission towards positive offsets that could be linked to a northern \hi\ asymmetry.

\subsection{Cold gas masses, fractions and deficiencies}\label{sec:gasfracs}

We present in Table \ref{table:logSmass} the molecular and atomic gas masses and deficiencies for the four galaxies in the sample. The full procedure for their estimation can be found in Apps. \ref{appendix:h2mass} and \ref{appendix:himass}. \hi\ and H$_2$ gas fractions quantify the remaining atomic and molecular gas relative to stellar mass, while \hi\ and H$_2$ deficiencies measure how depleted these reservoirs are compared to isolated galaxies of similar properties. Together, they provide complementary views of the state of environmental gas removal.

\begin{figure}
    \centering    \includegraphics[width=\columnwidth]{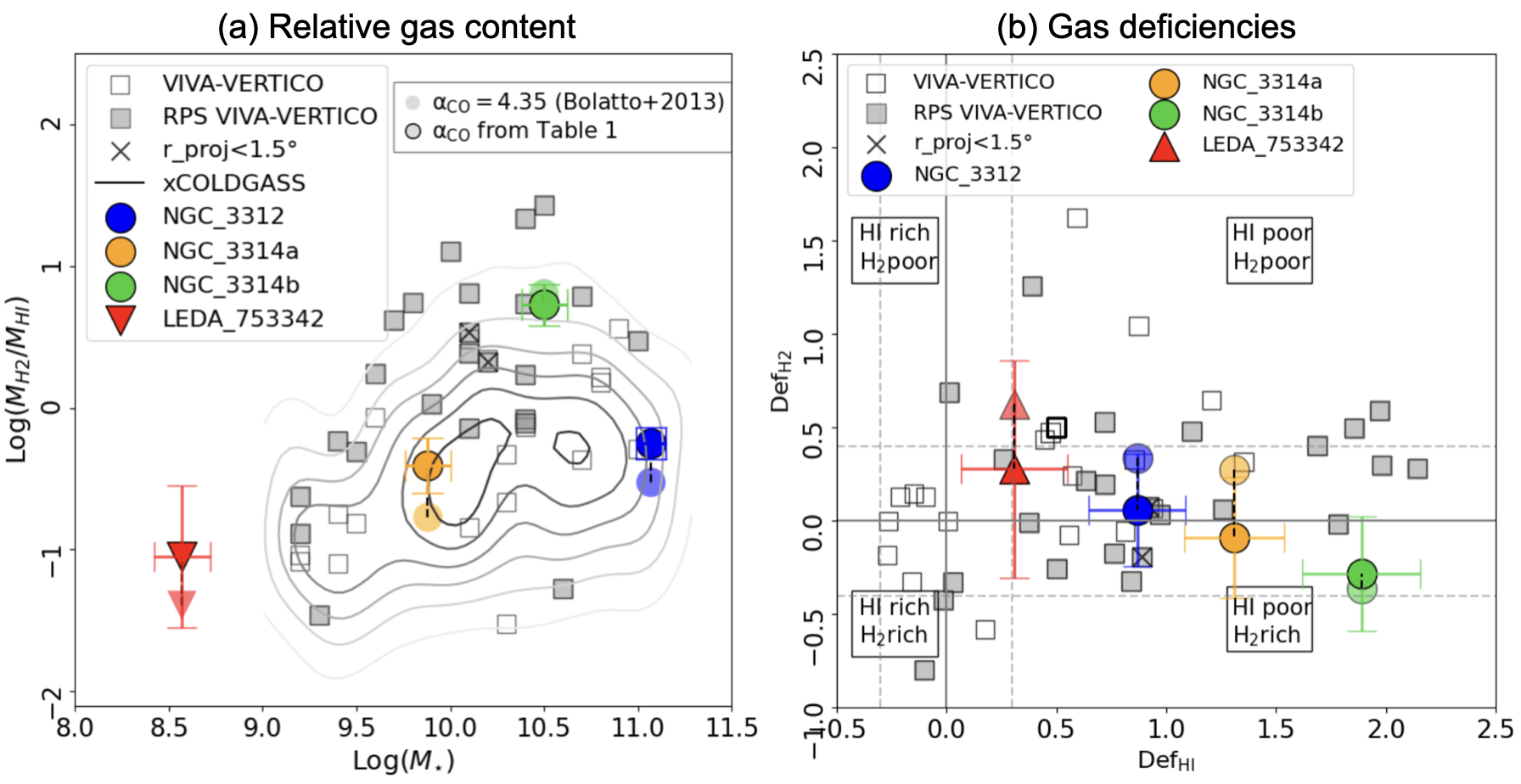}
    \caption{(a) H$_2$-to-\hi\ for galaxies NGC 3312, NGC 3314a/b and LEDA 753342. An upper limit is provided for LEDA 753342. For reference, xCOLD GASS galaxies are plotted as contours in the background and VIVA-VERTICO Virgo galaxies are grey squares. (b) \hi\ vs H$_2$ deficiency. NGC 3312, NGC 3314a/b and LEDA 753342 are shown in blue, yellow, green, red respectively. Dashed lines display the 0.3dex Def$_{\hi}$ \citep{2023Scholte} and the 0.4dex Def$\rm _{H_2}$ \citep{2013Scott_codex}.} \label{fig:coldgasfrac}
\end{figure}

Fig. \ref{fig:coldgasfrac} \textcolor{cyan}{a} displays the $\rm H_{2}$ to \hi\ ratio versus stellar mass for the galaxies in this work compared to xCOLDGASS field galaxies and cluster galaxies from Virgo. The xCOLDGASS sample \citep{saintonge2017xcold} provides the reference frame for galaxies with stellar mass between $\rm 10^9~M_{\odot}$ and $\rm 10^{11.5}~M_{\odot}$ and is displayed in gray in the background. Virgo galaxies from VIVA-VERTICO are distributed over a similar mass range, and are shown as filled grey squares if they are RPS according to \cite{yoon2017history}.  

 For comparison, we also calculate the gas fractions using the global $\alpha_{\rm CO}$ from \cite{2013Bolatto} instead of the stellar-mass-dependent $\alpha_{\rm CO}$, to demonstrate the sensitivity of our results to this choice of conversion factor. The constant conversion factor was used by \cite{saintonge2017xcold} to estimate M$\rm _{H_2}$ for the xCOLDGASS sample. The constant conversion factor predicts a higher relative amount of molecular gas for NGC 3312, NGC 3314a and NGC 3314b than the metallicity-dependent $\alpha_{\rm CO}$, while for LEDA 753342 it predicts a lower M$\rm_{H_2}$.

Figure \ref{fig:coldgasfrac} is adapted of Fig. 2 from \cite{Zabel2022}. It displays the gas deficiencies in the RPS galaxies from this work and the VIVA-VERTICO galaxies using the xCOLDGASS sample as a reference for the expected gas content, as explained in eq. \ref{eq:h2def}. The space is separated in four quadrants, depending on whether the galaxies are \hi\ rich or poor, and H$_2$ rich or poor. Most VIVA-VERTICO galaxies are in the \hi\ poor + H$_{2}$ poor quadrant. Dashed vertical lines display the \hi\ dispersion estimated by \cite{2023Scholte}, while dashed horizontal lines display the H$_2$ dispersion reported by \cite{2013Scott_codex}.

Galaxies NGC 3312 and NGC 3314a in Fig. \ref{fig:coldgasfrac} \textcolor{cyan}{a} follow the general trend displayed by the xCOLDGASS galaxies, but fall below most of the VIVA-VERTICO cluster galaxies with similar stellar masses. The galaxy NGC 3314b has a higher $\rm H_2/HI$ ratio relative to the general tendency for field galaxies (Fig. \ref{fig:coldgasfrac}), but falls within the average of Virgo galaxies. For LEDA 753342 we offer an upper limit to the molecular gas content. Its low metallicity implied using a larger conversion factor X$_{\rm CO}$ to accurately estimate this upper limit (see Table \ref{table:logSmass}).

Figure \ref{fig:coldgasfrac} \textcolor{cyan}{b} displays how the four galaxies are neither H$_2$ rich nor H$_2$ deficient. Within uncertainties, they trace a sequence of H$_2$ enrichment as the \hi\ depletion increases. LEDA 753342 is barely \hi\ and H$_2$ deficient. NGC 3312 has H$_2$ and \hi\ deficiencies compatible with that of Virgo RPS galaxies within 1.5$^{\circ}$ of the cluster core. 
NGC 3314a is \hi\ deficient, but its molecular gas content may not be noticeably affected. Finally, NGC 3314b is extremely \hi\ deficient and possibly slightly H$_2$ rich.

\section{Discussion} \label{sec:discussion}

\subsection{Compression in the gas-rich galaxy NGC 3312}

The asymmetric, \hi\ tailed morphology of NGC 3312 (Fig. \ref{fig:ngc3312}) is a direct result of the interaction between rotation and RPS. This effect has been reproduced in galaxy-scale wind-tunnel simulations \citep[e.g.][]{2001VollmerRPSsimu,2023Akkerman}. \cite{2025Souchereaujellyfish} discussed that RPS extracts gas from the side of the galaxy rotating with the wind, which is moved by angular momentum to the side rotating into the wind, where the tail forms from the concentrated gas and fallback may occur. The \hi\ tail in NGC 3312 extends to 40 kpc at column densities of N$_{\hi}\simeq10^{19}$ cm$^{-2}$: $\sim$10 kpc longer than initially reported by \citealt{Hess2022} with higher column densities of N$_{\hi}\simeq10^{20}$ cm$^{-2}$.   

Following \cite{Hess2022}, if we assume that the velocity of the intracluster medium is given by $v_{ICM} = \sqrt3\sigma_v$, where $\sigma_v=620$\kms\ is the velocity dispersion of the cluster \citep{lima2021environmental}, the resulting relative velocity of NGC 3312 and NGC 3314a with respect to the ICM is 1100\kms. We calculate a lower limit of the timescale of stripping, due to uncertainty in the projection and the assumption that gas would instantly stop with respect to the ICM. This limit is of $\rm \sim 35.6$ Myr, ten times lower than the timescales at which stripped gas remains neutral discussed by works such as \cite{2005Osterloo}.
We estimate that the total \hi\ mass stripped from NGC 3312 (see App. \ref{App1:optIm}) is M$_{\hi}\sim 4\times10^{8}$ \msun, around 8\% of the total \hi\ mass. Only 8\% of its \hi\ is outside the optical disk, similar to the quantity reported by \cite{Hess2022}. Thus, deeper imaging reveals a more extended \hi\ tail, though lower column density gas contributes negligibly to the total \hi\ mass.

 The high stellar content of NGC 3312 may be helping to retain atomic gas to a certain extent and causing a low \hi\ deficiency. This has also been observed by \cite{Luber2022extremeGASP} in 5 GASP galaxies with confirmed MUSE optical tails and \hi\ masses ranging  1-$7\times10^9$M$_\odot$ \citep[JO 135, JO 194, JO 201, JO 204, and JO 206 from][]{Poggianti2017}. Using VLA observations they reasoned, in agreement with previous results from \cite{Jaffe2018_GASP}, that the availability of more gas as a result of a bigger stellar content allowed their \hi\ reservoirs to survive for longer than those of lower mass cluster disk galaxies. A relatively recent infall or low inclination with respect to the ICM wind could also prevent extreme \hi\ deficiencies in NGC 3312.

The steep gradient in surface density displayed by the \hi\ contours on the eastern edge indicate that gas is compressed by the ram pressure. This effect has been observed in a number of targeted Virgo cluster galaxies undergoing RPS, either in the optical \citep[NGC 4402 and NGC 4522 from][]{abramson2016hst}, or through measurements of the molecular gas distribution \citep[examples are NGC 4921 or NGC 4330, NGC 4402 and NGC 4522 from][]{cramer2021molecular, lee2017effect}. The upper unwinding arm of NGC 3312 is bluer with a high concentration of dust slightly shifted to the west \citep{Hess2022}, and concentrates 33\% of CO and 18\% of the total H$\alpha$ emission. The leading arm of NGC 3312 is bright in the blue, contains a defined dust lane \cite{Hess2022}, 47\% of the total CO flux and 48\% of the total H$\alpha$ flux tracing star formation. Compression makes the galaxy subject to global starbursts, and in the case of NGC 3312 is responsible of the high concentration of star formation and new stellar populations, dust, and CO. We propose that the pressure from the ICM is what has triggered the formation of more CO in the leading and the unwinding arms. This could imply that RPS is increasing the global \hi-to-H$_2$ efficiency conversion. Rapid consumption following a compression-induced burst of star formation or gas redistribution could explain how the local increase in the \hi-to-H$_2$ conversion in the leading arm can coexist with a moderate overall molecular gas deficiency (see Fig. \ref{fig:coldgasfrac}).

\subsection{The jellyfish galaxy NGC 3314a} \label{sec:3314a_discussion}

NGC 3314a (Fig. \ref{fig:ngc3314a}) is a spiral galaxy with 
centrally concentrated CO, found in regions with \hi\ levels above $\rm N_{HI}\geq10^{20}~cm^{-2}$. Its $\rm M_{H_2}/M_{HI}$ content is within the scatter expected from field galaxies. The decrease in H$_2$ deficiency relative to NGC 3312 could be due to the same process of compression proposed for this galaxy. The elevated \hi\ deficiency (Fig. \ref{fig:coldgasfrac}) could be a combination of enhanced \hi-to-H$_2$ conversion and RPS. 

NGC 3314a displays long tails of neutral atomic gas whose kinematics mirror that of the disk. Their length is between $\sim 50$ kpc and $\sim 70$ kpc in projection (see Sec. \ref{sec:3314a_results}). The first measurement is fully consistent with the length estimated by \cite{Iodice2021_ngc3314} with VST data, and the second almost doubles the previous estimated extent of the stripped material \citep{Hess2022}. This length is typical of jellyfish galaxies, extreme cases of RPS in which the gas that is removed from the galaxy forms long tails of material capable of sustaining star formation \citep{kenney2013transformation,gullieuszik2020gasp} which may give rise to dark matter-free ultra-diffuse galaxies \citep{poggianti2019gasp}. We calculate a lower limit of neutral gas of M$_{\hi}=6\times10^{8}$M$_{\odot}$ inside the tail (see App. \ref{App1:optIm}), within column densities of N$_{\hi}\sim 5\times 10^{18} \rm cm^{-2}$. This is approximately 33\% of the total \hi\ content, in consonance with the 35\% measured by \cite{Hess2022}.  

Assuming that the stripped gas is instantly slowed down to ICM rest velocity, and that NGC 3314 moves at a relative velocity of $\sim$1100\kms with respect to the ICM, we estimate a lower limit timescale for stripping of $\sim45$~Myr. \cite{Hess2022} applied the same method and gave a conservative range for the stripping timescale of 33-53 Myr. \cite{Hess2022} discussed that traditional analytic models have typically estimated timescales of the order of 100 Myr of \citep{schulz2001multi, roediger2005ram}, considering corrections for projection or gas deceleration. However, more recent simulations propose that these timescales underpredict the stripping rate of neutral gas and overpredict that of ionized gas due to an underestimation of stellar feedback that enhances RPS, gas compaction that suppresses it \citep[e.g.][]{Kulier2023}, and radiative cooling or magnetic fields \citep{Tonnesen2019}. These simulations suggest that the shorter timescale we estimate may be more realistic.

We have observed that the southeastern tail could be host to the predecessor of a dwarf galaxy born in situ from RPS. This SF+\hi\ clump is more similar to a tidal dwarf galaxy than to the clumps of star formation inhabiting stripped gas, that have median stellar masses of $\sim10^6$ M$_{\odot}$ and median radius of $<200$ pc \citep{poggianti2019gasp}. Although we cannot establish its velocity dispersion and clarify if the overdensity could be self-sustained by gravitation, it could be a candidate for predecessor of a dark-matter-free dwarf galaxy born from the collapse of RPS material. The idea of a RPS dark-matter free dwarf galaxy has been proposed in other extremely RPS galaxies such as ESO 137-001 \citep{jachym2014abundant}. Simulations \citep[e.g. ][ from IllustrisTNG-50]{2024Lora} have also confirmed that self-bound, dark-matter free star formation regions of gas and stars of masses $\sim2\times10^{8}$M$_{\odot}$ can be formed in RPS tails. 

The south-western tail is perhaps the progenitor of UDG32. This UDG was first reported by \cite{iodice2021formation}, who proposed it could have formed from RPS, and \cite{hartke2025looking} studied it as a promising case of UDG formed by RPS.  While \cite{hartke2025looking} measured a line-of-sight velocity of UDG32 that agrees with that of the H$\alpha$ filament, they argue its high metallicity and intermediate age is at odds with a recent formation from RPS material and in favor of a pre-existing UDG potentially interacting with the \hi\ tail. However, they do not rule out a formation from pre-enriched material. \cite{hartke2025looking} estimated that the line of sight velocity for UDG32 ($v_{UDG32}=3085 \pm 120 \rm km/s$), which is similar enough to the velocity of the \hi\ co-spatial overdensity ($v_{tail}=3033\pm67 \rm km/s$) to be an additional observational constraint in favor of the RPS scenario. 

\subsection{The gas deficiency of galaxy NGC 3314b}

NGC 3314b (Fig. \ref{fig:ngc3314b}) is the most advanced stripping case in the group. It presents an extremely deficient and truncated \hi\ disk. Assuming edge-on stripping, the \hi\ spatial asymmetries seen at the highest spatial resolution (Figs. \ref{fig:ngc3314b} \textcolor{cyan}{b}, \textcolor{cyan}{d}), in combination with its elevated \hi\ deficiency (Fig. \ref{fig:coldgasfrac} \textcolor{cyan}{b}), could suggest the presence of a mostly evaporated neutral gas tail further northeast, beyond the stellar disk. The stripped \hi might be starting to leave the central molecular gas less shielded against the direct influence of the ICM. Its M$\rm_{H{_2}}$/M$_{\hi}$ fraction and H$_2$ richness are the most elevated in the group. The increased pressure caused by ram pressure from the leading northwestern side may be promoting \hi-to-H$_2$ conversion, as reported in Virgo galaxies by \cite{roberts2023vertico}, and be responsible for the H$_2$ asymmetrical distribution, shifted to the northwest (Fig. \ref{fig:ngc3314b}). These results are often a product of extended exposure to the ICM \citep{Vollmer12,boselli2014IIIcold,Cybulski2016,2021Loni}, which supports a previous observation by \cite{Hess2022} that NGC 3314b might be the oldest cluster member among the four galaxies. This galaxy may serve as a reference for what the future of NGC 3312 and NGC 3314a gas contents will look like. 

\subsection{Insights on the origin of LEDA 753342} \label{sect:insights}

LEDA 753342 (Fig. \ref{fig:leda753342}) is a ring galaxy with no detected CO. The residual light to the south of this galaxy (Fig. \ref{fig:leda753342} \textcolor{cyan}{a}) may originate in a past gravitational interaction that triggered annular shocks of star formation in its disk, and perhaps a northern \hi\ feature that is not completely resolved by our high resolution  \hi\ observations (16$^{\prime\prime}$x11$^{\prime\prime}$). Similar features have been reported for ring galaxies in works such as \cite{Timmis2017_rings}. 

The northern \hi\ asymmetry (Fig. \ref{fig:leda753342} \textcolor{cyan}{a}) and slight \hi\ deficiency (Fig. \ref{fig:coldgasfrac} \textcolor{cyan}{b}) are compatible with a recent RPS scenario, since low mass galaxies lose gas very rapidly in the cluster environment \citep{Serra23_Fornax_HI_survey}, particularly when transiting through the cluster core. LEDA 753342 is slightly \hi\ richer than the six Fornax dwarf galaxies undergoing RPS presented by \cite{zabel2024resolved}, and will likely follow their evolution as RPS disturbs its gas. As proposed for the blue irregular galaxy NGC 1427A in the Fornax cluster \cite{2024Serra}, LEDA 753342 could have suffered a previous gravitational interaction, yet its \hi\ distribution and kinematics may be shaped by ram pressure.  

\subsection{Cold gas stripping and its impact on star formation quenching}

We discussed in previous sections the cold gas content, its distribution and deficiency for galaxies NGC 3312, NGC 3314a/b and LEDA 753342. 
We focused on RPS as the primary driver of gas removal, although recent work in lower-mass clusters such as Fornax \citep{2018LeeWaddell,2021Loni,ramatsoku2025meerkat} shows that tidal interactions can rival or even exceed RPS. Our selected galaxies are located close to the X-ray peak of Hydra I \citep[see e.g.][]{2026Srivastava}, within the stripping and virialized regions in phase space \citep[e.g.][]{delaCasa2025}, so the impact of RPS is not negligible with respect to tidal interactions. The strength of RPS at a given projected distance to the cluster center also grows with cluster mass. A galaxy such as NGC 3312 ($\rm r_{proj} = 110~kpc, v_{rel}=865~km/s$) would experience a ram pressure strength given by P$_{\rm ram} = \rho_{\rm ICM}v_{\rm gal}^2 $. In the Virgo and Fornax clusters this would amount to $\rm P_{Virgo} = 2.1\times10^{-11} ~g~cm^{-1}~s^{-2}$, $\rm P_{Hydra} = 3.9\times10^{-12}~g~cm^{-1}~s^{-2}$ and $\rm P_{Fornax} = 6.2\times10^{-13}~g~cm^{-1}~s^{-2}$ (see App. \ref{app:rps_strength} for details on the estimation of $\rm \rho_{ICM}$). Virgo would display the strongest ram pressures, followed by Hydra I, and Fornax, with a difference of an order of magnitude between each cluster. The fact that the galaxies in this work have relative cold gas fractions and deficiencies between Virgo and xCOLDGASS galaxies may be related to Virgo having a halo mass twice as massive \citep[6.3$\pm$0.9$\times$10$^{14}$M$_{\odot}$, ][]{2020Kashibadze} as Hydra I ($3.02\times10^{14}$M$_{\odot}$, see Sec. \ref{sec:Intro}) leading to a ram pressure ten times stronger.

The H$_{2}$/\hi\ fraction and deficiencies for the four galaxies range between the field and Virgo distributions, spanning different stages of RPS. Despite the long tails of NGC 3312 and NGC 3314a, no considerable amount of gas is lost for these galaxies. RPS could be promoting star formation via gas compression in NGC 3312, by enhancing \hi-to-H$_2$ conversion and H$_2$ consumption. This galaxy is consistent with an early stripping phase triggering molecular gas formation and redistribution followed by enhanced star formation. NGC 3314a represents an intermediate RPS stage characterized by a long gaseous tail and molecular gas largely confined to the central regions. 
NGC 3314b in contrast, is in a late stripping phase with a \hi\ deficient disk that is leaving its asymmetric molecular reservoir exposed. Finally, the low-mass LEDA 753342 is possibly influenced by a combination of a past tidal interaction and an early-stage RPS. Together, these galaxies trace a coherent evolutionary sequence of gas removal and transformation within the Hydra I cluster.

\section{Summary} \label{sec:summary}

We present in this paper deep CO and \hi\ observations in four Hydra I galaxies allowing the study of this cluster with the same multiphase perspective (optical-HI-CO) that already exists for other clusters such as Virgo and Fornax. The deep CO observations have similar spatial resolution as the deep \hi\ data presented in this work. We use them to compare the H$_2$ and \hi\ gas of four galaxies in the active stripping phase and relatively high star formation rates. In contrast to previous works in the literature performed on more heterogeneous samples, we analyze the morphology, gas fractions and deficiencies of galaxies within the physically related SE substructure in Hydra I proposed by \cite{Spavone2024_HydraStructure}, and linked them to their RPS stage.  

 NGC 3312 is a \hi-rich galaxy with a global burst of star formation \citep{delaCasa2025} that is related to RPS increasing the pressure in its leading edge. NGC 3314a is a spiral galaxy suffering edge-on RPS in a more impactful way for its gas deficiency and morphology. It has a long \hi\ tail with no detected CO that could be host to a newborn dwarf galaxy and UDG32. NGC 3314b is in a more advanced stage of RPS, having lost a significant amount of neutral gas that is leaving the molecular gas reservoir exposed. Finally, LEDA 753342 is a low mass galaxy with no detected molecular gas, that presents stellar and \hi\ disk asymmetries that could be linked to a past gravitational interaction combined with recent RPS.

 None of the studied tails have detectable levels of CO in our observations. We detect CO within the stellar disks at column densities of $\rm N_{\hi}\sim 10^{20} cm^{-2}$, which is disturbed as a product of RPS. In the case of the starburst galaxy NGC 3312, we observe that the highest CO contents are located in the windward half and unwinding spiral arms, which also present high H$\alpha$ fluxes. We reason that molecular gas disturbances may be connected to the relatively high star formation rates of this galaxy. Thus the effect of RPS on star formation quenching may be twofold: it reduces the amount of material available to form stars, and it may promote the consumption of the remaining gas that leads to global starbursts in the short-term, and faster quenching in the long-term. Future work involving the study of molecular gas consumption timescales will test this hypothesis.

The H$_2$/\hi\ ratio of NGC 3312, and NGC 3314a/b is within the VIVA-VERTICO galaxies distribution, but in the upper end of the xCOLDGASS galaxies. The behavior of LEDA 753342 is in agreement with that of other dwarf galaxies in Fornax \citep[e.g.][]{2023Kleiner_fornaxdwarf,2024Serra}. They manifest an intermediate stage between field and more virialized cluster galaxies. 

In conclusion, these four Hydra I galaxies reflect different stages along the stripping sequence. They have \hi\ tails, except perhaps LEDA 753342, disturbed H$_2$ disks and altered gas ratios that place them between field galaxies and more processed populations of more massive clusters. NGC 3312, NGC 3314a and LEDA 753342 may be an early instance of \hi\ gas removal and promoted transformation into H$_2$ by RPS, while NGC 3314b displays the likely future of these galaxies. They are proof of the dynamic stage of assembly of the Hydra I cluster, both as individual galaxies and as potential members of a substructure in the case of NGC 3312 and NGC 3314a. Recent infall into Hydra I may have triggered rapid removal of \hi\ gas and short-lived increase in molecular gas formation in regions exposed to the highest ICM wind pressures. This phenomena could be related to brief global increases in star formation efficiency, according to the observations from \cite{delaCasa2025}, setting RPS galaxies on an accelerated evolutionary path toward quenching.

\begin{acknowledgements}\\

     The authors kindly thank the referee for their insightful remarks and major contributions to improve this manuscript. We also thank Paolo Serra and Carmen Toribio for revising the manuscript and helping to improve it. This paper makes use of the following ALMA data: 2022.1.00716.S. This publication has received funding from the European Union’s Horizon 2020 research and innovation programme under grant agreement No 101004719 (ORP). CCC acknowledges support from the Nordic ALMA Regional Centre (ARC) node based at Onsala Space Observatory. The Nordic ARC node is funded through Swedish Research Council grant No 2019-00208. CCC and KH acknowledge support from the Sweden SKA Regional Center (sweSRC) node operated by Onsala Space Observatory in collaboration with Chalmers e-Commons. The Onsala Space Observatory national research infrastructure is funded through Swedish Research Council grant No 2019-00208. CCC, LV, BN, AS, RI acknowledge financial support from the grant CEX2021-001131-S funded by MICIU/AEI/ 10.13039/501100011033, from the grant PID2021-123930OB-C21 and PID2024-155817OB-I00 funded by MICIU/AEI/ 10.13039/50110001103x3 and by ERDF/EU. BN also acknowledges the grant DGP\_POST\_2024\_01021 funded by la Junta de Andalucía /CUII and by the ESF+. NZ is supported through the South African Research Chairs Initiative of the Department of Science and Technology and National Research Foundation. YLJ  acknowledge support from the Agencia Nacional de Investigaci\'on y Desarrollo (ANID) through Basal project FB210003, FONDECYT Regular projects 1241426 and 123044, and  Millennium Science Initiative Program NCN2024\_112. JH acknowledges support from the UK SKA Regional Centre (UKSRC). The UKSRC is a collaboration between the University of Cambridge, University of Edinburgh, Durham University, University of Hertfordshire, University of Manchester, University College London, and the UKRI Science and Technology Facilities Council (STFC) Scientific Computing at RAL. The UKSRC is supported by funding from the UKRI STFC. We acknowledge the use of the ilifu cloud computing facility – www.ilifu.ac.za, a partnership between the University of Cape Town, the University of the Western Cape, Stellenbosch University, Sol Plaatje University and the Cape Peninsula University of Technology. The ilifu facility is supported by contributions from the Inter-University Institute for Data Intensive Astronomy (IDIA – a partnership between the University of Cape Town, the University of Pretoria and the University of the Western Cape), the Computational Biology division at UCT and the Data Intensive Research Initiative of South Africa (DIRISA). The MeerKAT telescope is operated by the South African Radio Astronomy Observatory, which is a facility of the National Research Foundation, an agency of the Department of Science and Innovation. This research made use of Photutils, an Astropy package for detection and photometry of astronomical sources \citep{larrybradley2024}.

\end{acknowledgements}

\bibliographystyle{aa}
\bibliography{library.bib}

\begin{appendix} 

\section{$\rm H_2$ masses and molecular gas deficiencies from CO fluxes}\label{appendix:h2mass}

We used the standard equation (1) to estimate the $\rm M_{H_2}$ from the total flux of the line observed $\rm \int S_{\nu}d\nu$:

\begin{equation} \label{eq:mh2}
   \rm M_{H_2} = 2m_HD^2X_{CO}\frac{\lambda^2}{2k_B}\int S_{\nu}d\nu
\end{equation}

To estimate $\rm \int S_{\nu}d\nu$ for NGC 3312, NGC 3314a and NGC 3314b, we integrated the CO(1-0) emission inside the source mask defined by the SoFiA source finder. To obtain an upper limit of the flux for LEDA 753342, which had no associated detection, we estimated the size of the galaxy in number of resolved elements, and factored it by the RMS noise and the line width to get a minimum flux. A full description of this process can be found in App. \ref{app:nondets}.

The rest of parameters in eq. \ref{eq:mh2} are the mass of a hydrogen atom $\rm m_H$, the distance to the galaxy $\rm D$, the rest wavelength of the line observed $\lambda$, the Boltzmann constant $\rm k_B$, and the CO-to-H$_2$ mass conversion factor $\rm X_{CO}$. To calculate $\rm X_{CO}$ in units of [$\rm M_{\odot}pc^{-2}(Kkms^{-1})^{-1}$] for each galaxy, we used the standard relation with the metallicity-dependent conversion factor $\alpha_{\rm CO}$:

\begin{equation}
   \rm X_{CO} = 2.12\times10^{20} \alpha_{CO}
\end{equation}

where $\alpha_{C0}$ was obtained following eq. (25) from \cite{accurso2017deriving} and can be found in Table \ref{table:logSmass}:

\begin{equation}
    \rm log\alpha_{CO} = 14.752-1.623 \left[12+log\left(O/H\right)\right] +0.062 log \left(\Delta MS\right)
\end{equation}

being $\rm \left[12+log\left(O/H\right)\right]$ the global metallicity of the galaxy that we estimated following eq. (1) from \cite{Sanchez2017}, which parameterizes metallicity in terms of the stellar content M$_{\star}$: 

\begin{equation}
\begin{split}
    \rm \left[12+log\left(O/H\right)\right] = \\ \rm  8.74  
    +
    0.005\times \left[log\left( \frac{M_{\star}}{M_{\odot}}\right)-8-3.5\right]\times exp \left[3.5-log\left( \frac{M_{\star}}{M_{\odot}}\right)-8\right]
\end{split}
\end{equation} 

Finally, we took the stellar masses M$_{\star}$ derived from WISE \citep[Wide-field Infrared Survey Explorer ,][]{Jarrett2019} and H$\alpha$-derived distance to the main sequence $\rm \Delta MS$ values from \cite{delaCasa2025}, as explained in sec. \ref{sec:DECam_data}. They are displayed in Table \ref{table:logSmass} with their uncertainties, that are a result of propagating the errors of the integrated observed line, the dispersion of the metallicity parameterization \citep[0.06dex,][]{Sanchez2017}, and the 1$\sigma$ spread of $\rm log\alpha_{CO}$ \citep[0.16dex,][]{accurso2017deriving}.

$\rm H_{2}$ deficiencies were estimated by subtracting our observed $\rm H_2$ masses from those expected from the xCOLDGASS sample:

\begin{equation}\label{eq:h2def}
    \rm DEF_{H_2} = log_{10}(M_{xCOLDGASS}/M_{\odot}) - log_{10}(M_{obs}/M_{\odot})
\end{equation}

To estimate the expected $\rm log_{10}(M_{xCOLDGASS}/M_{\odot})$ for each mass we followed the procedure in \cite{zabel2019alma}. We separated the xCOLDGASS sample in 6 equidistant mass bins in the range $\rm 9<\rm log(M_{\star}/M_{\odot})<12.5$. This number of bins ensured that there were $~50$ galaxies per bin. We took the median value of the $\rm log_{10}(M_{xCOLDGASS}/M_{\odot})$ and $\rm log_{10}(M_{\star}/M_{\odot})$ as reference, and then interpolated to obtain the expected molecular masses in intermediate stellar masses. The deficiencies are presented in Table \ref{table:logSmass}.

\section{\hi\ masses and atomic gas deficiencies}\label{appendix:himass}

\hi\ masses were calculated through equation (48) from \cite{meyer2017tracing}:

\begin{equation}
    \rm \left(\frac{M_{\hi}}{h_{C}^{-2}M_{\odot}}\right) \simeq 49.7 \left(\frac{D_L}{h_{C}^{-2}Mpc}\right)^{2}\left(\frac{S}{JyHz}\right)
\end{equation}

where $\rm D_L$ is the luminosity distance to the galaxy, and S is the observed flux. Uncertainties were propagated from the error of the integrated line.

\hi\ deficiencies $\rm DEF_{\hi}$ were estimated as: 

\begin{equation}\label{eq:hIdef}
    \rm DEF_{\hi} = log_{10}(M_{exp}/M_{\odot}) - log_{10}(M_{obs}/M_{\odot})
\end{equation}

Where $\rm M_{obs}$ is the observed mass, and $\rm M_{exp}$ is the predicted \hi\ mass. As a predictor, we use the AMIGA sample of isolated galaxies, that provides an evolutionary reference for galaxies in isolation. We estimate the expected mass from the scalar relation between M$\rm _{HI}$ and the optical luminosity $\rm L_B$ presented by \cite{jones2018amiga}:

\begin{equation}\label{eq:hIdef}
    \rm log(M_{exp}/M_{\odot}) = 0.94~log(L_B/L_{\odot}) + 0.18 
\end{equation}

We convert DECam magnitudes reported in \cite{delaCasa2025} from the \textit{ugriz} system to the UBVRI system using the empirical color transformations from \cite{jordi2006_ugriz}. This provides an accurate estimate for galaxies with stellar masses M$_{\star}/$M$_{\odot}<10^{9.2}$, which is the approximate lower stellar mass limit of the AMIGA sample. For LEDA 753342, whose stellar mass is $\rm log(M_{\star}/M_{\odot})=8.57$, we adopt the same approach as \cite{sorgho2024amiga}. They estimated the following relation for galaxies with M$_{\star}/$M$_{\odot}<10^{9.2}$ from low mass galaxies in the NASA Sloan Atlas catalog \citep{2011Blanton}. For galaxies with stellar masses M$_{\star}/$M$_{\odot}<10^{8.6}$, the expected \hi\ mass is obtained from:

\begin{equation}\label{eq:hIdef}
    \rm log(M_{exp}/M_{\odot}) = 0.46 log (M_{\star}/M_{\odot}) + 5.18 
\end{equation}

\section{CO flux in the spiral arms of NGC 3312} \label{App:ngc_3312}

\begin{figure}[h]
\includegraphics[width=9cm]{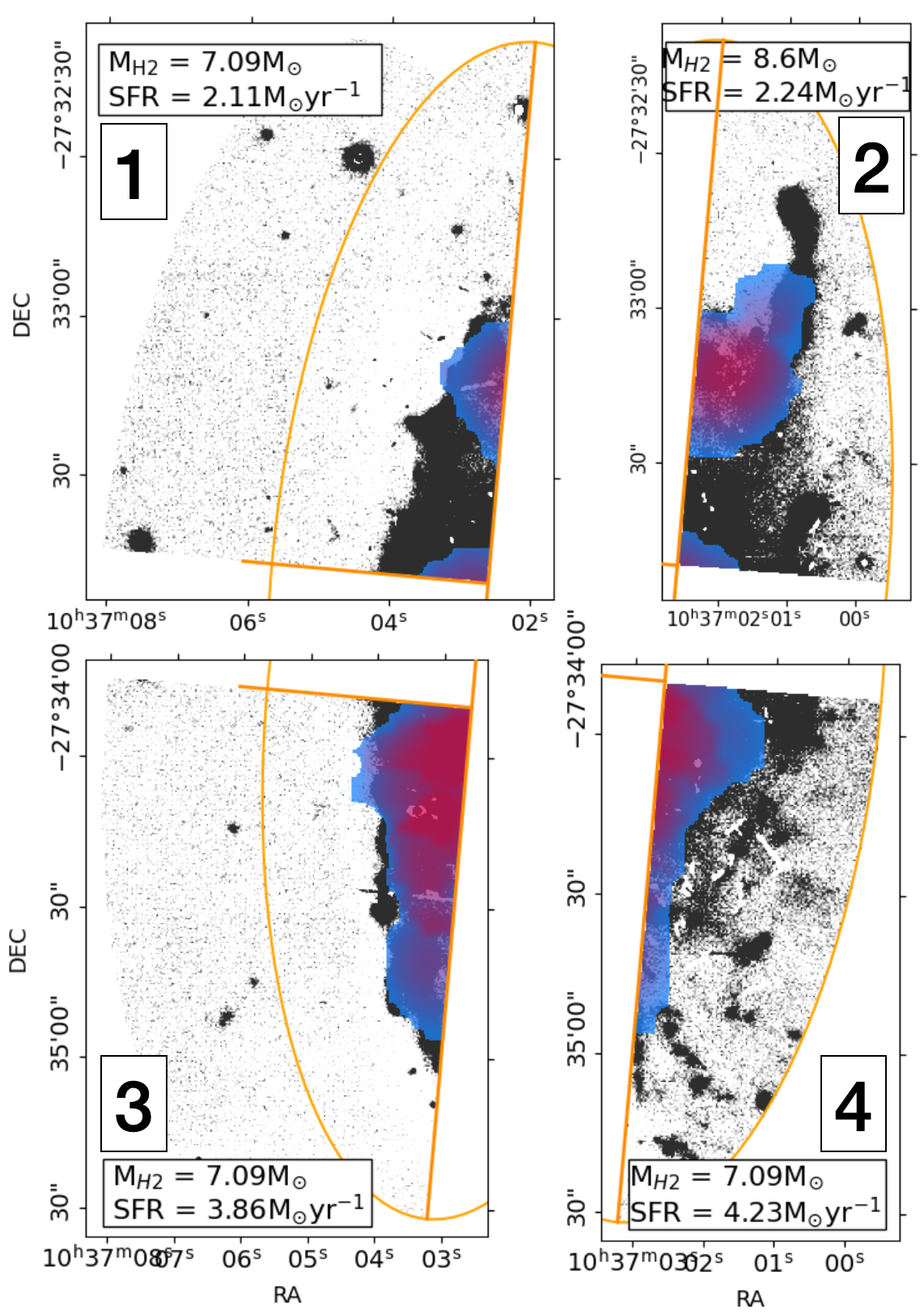}
\centering
 \caption{ DECam H$\alpha$ map of galaxy NGC 3312 with the r26 isophote (orange) and the CO moment 0 map (red-blue gradient) with column densities higher than N$_{\hi}=10^{20}$cm$^{-2}$. Each panel shows a different quadrant of the galaxy. Panel 1 (top left) is the windward upper quarter, panel 2 (top right) is the leeward upper quarter containing the unwinding spiral arm, panel 3 (bottom left) contains the windward lower quarter with most of the leading compressed spiral arm, and panel 4 (bottom right) is the downwind lower side of the galaxy with part of the tail.}
    \label{fig:app_ngc3312}
\end{figure}

To estimate the amount of CO concentrated in the leading and trailing side of the galaxy NGC 3312, we fitted the isophote at which the emission in the DECam r band decreases to the magnitude 26 (r26). We defined the leading and the trailing half by separating the ellipse in two through the semimajor axis. Then we separate again to isolate the unwinding arm, following the pattern displayed in Fig. \ref{fig:app_ngc3312}. We summed the CO flux in each region and estimated the corresponding H$_2$ mass applying Eq. \ref{eq:mh2}. The leading half concentrates 47\% of the total CO flux, while the trailing half has the remaining 53\%. We find that the majority of the CO flux is found in the lower right quadrant (40\%) containing the leading compressed spiral arm, and the upper right quadrant (33\%) hosting the unwinding a spiral arm. As for the percentage from the total H$\alpha$ flux, the region within the r26 isophote for panel 1 contains 17\%, 18\% for panel 2, 31\% for panel 3, and 34\% for panel 4. High CO fluxes seem to be followed by high H$\alpha$ flux, with the exception of panel 4 which includes H$\alpha$ emission from the tail.

\section{F$_{CO}$ estimation in regions without detections}\label{app:nondets}

\begin{figure}[t]
\includegraphics[width=8cm]{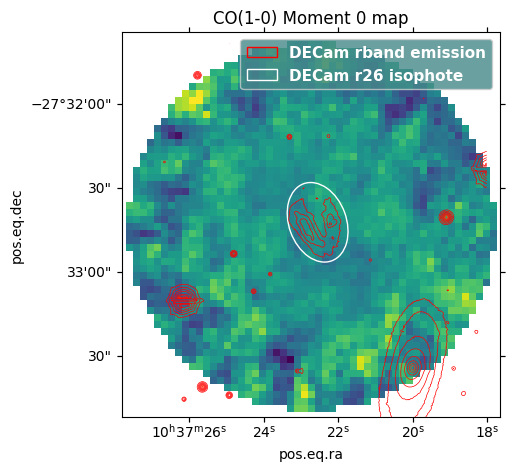}
\centering
 \caption{CO(1-0) moment 0 intensity map for LEDA 753342, from a cube centered at velocity $\rm v_{HI}=2748.704 km/s$ and a total width of 400 km/s. The DECam r26 isophote within which the signal was integrated is represented in white. DECam \textit{r} band contours are also shown as reference for the light distribution of the galaxy.}
    \label{fig:app_conondets_leda}
\end{figure}

We provide a generous upper limit for the molecular gas content within LEDA 753342 in Tab. \ref{table:logSmass}, since no detection was found by the sourcefinder. To estimate the flux F$\rm _{CO}$, we estimated the 3$\sigma$ noise level inside the 26 magnitude isophote in the \textit{r} band of the galaxy within noise-flattened cubes, and multiplied it by the number of resolution elements in the beam and the linewidth of the measured line in the galaxy:

\begin{equation}
    F_{CO} = 3\sigma \times \frac{pxl_{aperture}}{pxl_{beam}} \times w_{CO(1-0)}
\end{equation}

Where $w_{CO(1-0)} = 150 km/s$ is the linewidth of the \hi\ emission of LEDA753342, and $\frac{pxl_{aperture}}{pxl_{beam}} = 0.36$ is the number of resolution elements for the chosen aperture. We assumed a gaussian distribution of noise to calculate the RMS from the mean average deviation (MAD) SNR:    $\sigma = \sqrt{\left(\frac{2}{\pi}\right)}\times MAD $

The mass was estimated by replacing $\int S_{\nu}d\nu$ by $F_{CO}$ in Eq. \ref{eq:mh2}.

\begin{figure}[t]
\includegraphics[width=8cm]{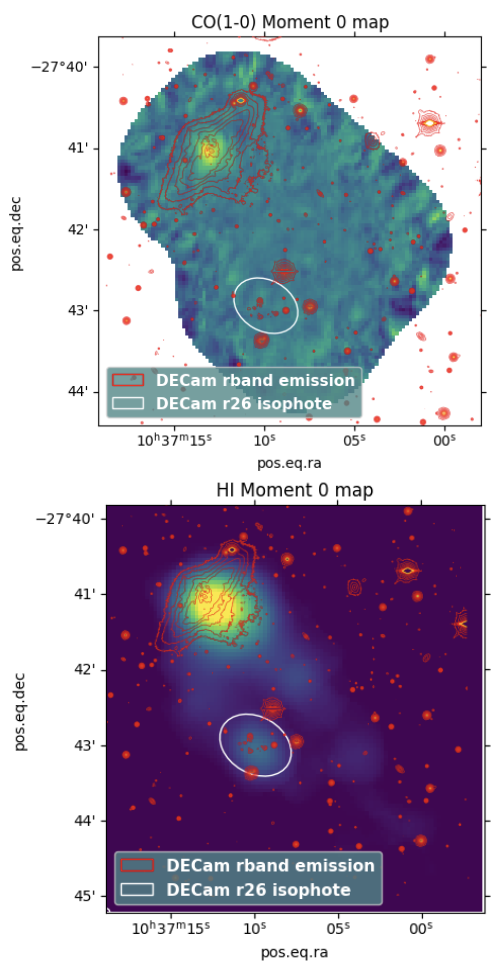}
\centering
 \caption{CO(1-0) moment 0 intensity map (top) and \hi\ moment 0 map (bottom) for the bright star-forming knot in the tail of NGC 3314a. The DECam r26 isophote within which the signal was integrated is represented in white. DECam \textit{r} band contours are also shown as reference for the light distribution of the galaxy.}
    \label{fig:app_conondets_3314a}
\end{figure}

The procedure to estimate the upper limit of molecular gas in the bright knot of star formation in the tail of NGC 3314a was identical to that of LEDA 753342. We considered a linewidth of $w_{CO(1-0)} = 125$ km/s, based on the \hi\ profile in the region. For the upper limit in \hi\, we  integrated the emission within the r26 isophote aperture in the \hi\ moment 0 map provided by the SoFiA sourcefinder, as displayed in Fig. \ref{fig:app_conondets_3314a}.

\section{\hi\ mass in tails} \label{App1:optIm}

\begin{figure}[h]
\includegraphics[width=8cm]{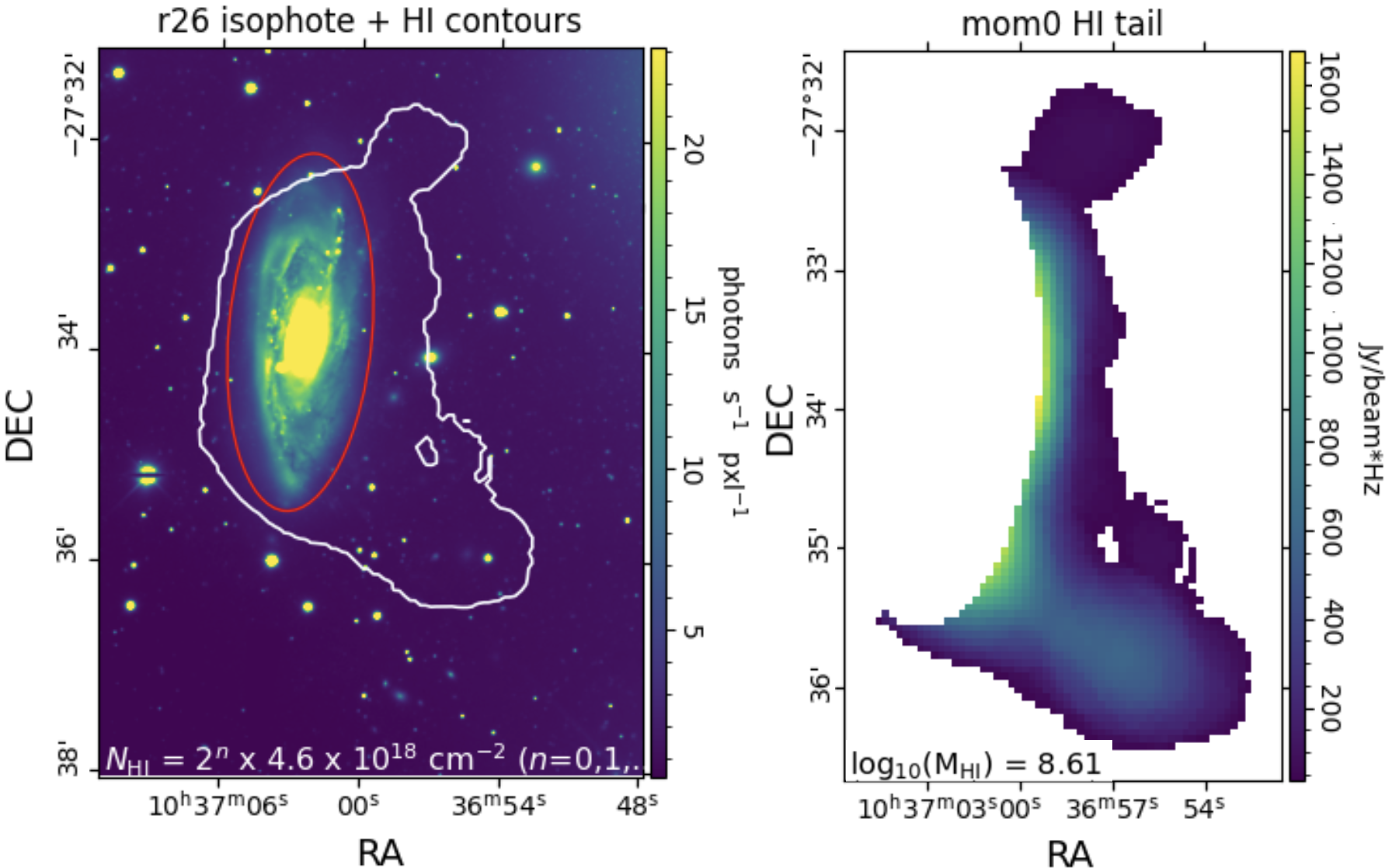}
\centering
 \caption{\textit{Left}: DECam \textit{r} band with \hi\ 45$^{\prime\prime}$x45$^{\prime\prime}$ beam contours of galaxy NGC 3312. The red ellipse represents the r26 isophote and is considered as a conservative limit to the extent of the optical disk. \textit{Right}: \hi\ moment map of the \hi\ tail without the galaxy disk. The \hi\ mass contained is in the bottom left corner. }
    \label{fig:app_tails1}
\end{figure}

\begin{figure}[h]
\includegraphics[width=9cm]{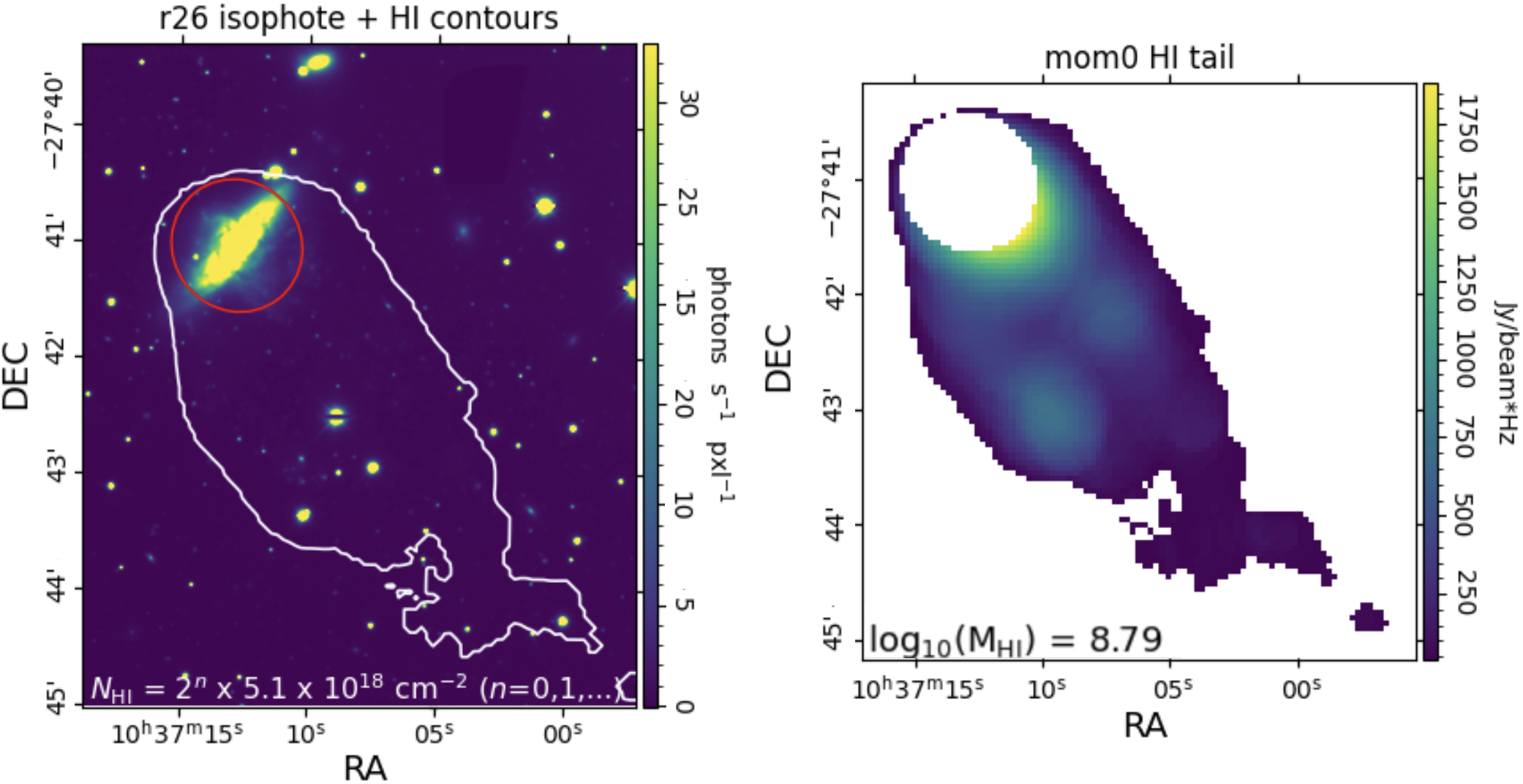}
\centering
 \caption{\textit{Left}: DECam \textit{r} band with \hi\ 45$^{\prime\prime}$x45$^{\prime\prime}$ beam contours of galaxy NGC 3314a. The red ellipse represents the r26 isophote and is considered as a conservative limit to the extent of the optical disk. \textit{Right}: \hi\ moment map of the \hi\ tail without the galaxy disk. The \hi\ mass contained is in the bottom left corner. }
    \label{fig:app_tails2}
\end{figure}

In Figs. \ref{fig:app_tails1} and \ref{fig:app_tails2} we show the r26 isophote for both galaxies over the \textit{r} band image, the masked \hi\ moment 0 intensity map, and the masked H$\alpha$ image, from left to right.

\section{Estimation of the ram pressure strength in different clusters}\label{app:rps_strength}

To make an estimation of the relative strength of ram pressure in the Hydra I cluster and the two reference clusters Virgo and Fornax, we computed the following general formula:

\begin{equation} \label{eq:Pram}
P_{\rm ram} = \rho_{\rm ICM}v_{\rm gal}^2 
\end{equation}

where $v_{\rm gal}$ is the relative velocity between a galaxy and the cluster core the ICM gas density profile follows a $\beta$-model: $\rho_{\rm ICM}=\rho_0[1+(\frac{r\pi}{2R_c})^2]^{-3\beta/2}$

For consistency, we take the values of the $\beta$-model parameter of the ICM gas density profile ($\beta$), the central initial density of the ICM ($\rho_0$), and the core radius (r$_c$) from the work of \cite{2007Chen_ppsdparams}. They can be found in Table \ref{table:RPS_force}.

\begin{center}
    \begin{table}[h!]
    \centering
    \caption{\label{table:RPS_force}Parameters used to estimate $\rm \rho_{ICM}$}
    \begin{tabular}{ c c c c} 
    \hline
       Parameter & Virgo &  Hydra I & Fornax \\

     \hline
       $\beta$ & 0.690 & 0.607 & 0.804 \\
       $\rm \rho_0 (cm^{-3})$ & 2.12$\times 10^{-2}$ & 4.7$\times 10^{-3}$ & 9$\times10^{-4}$ \\
       r$_c$(kpc) & 162 & 94 & 173 \\
       
     \hline
    \end{tabular}\\
    {\textit{Column 1:} Name of the parameter. $\beta$ is the $\beta$-model parameter of the ICM gas density profile, $\rho_0$ is the central initial density of the ICM, and r$_c$ is the core radius as fitted by \cite{2007Chen_ppsdparams}. 
     
    \textit{Column 2:} Values for the Virgo cluster. \textit{Column 3:} Values for the Hydra I cluster. \textit{Column 4:} Values for the Fornax cluster.}
    \end{table}
\end{center}

We select the galaxy NGC 3312, which is the closest in projection to the Hydra I cluster core (1112kpc) and has a relative velocity of v$_{gal}$ = 865km/s. A galaxy falling with the same relative velocity and at the same cluster-centric distance in different environments will experience different ram pressure forces due to variations in the intracluster medium density. Since the ICM is ionized, we assume that it contains mainly protons of mass $\rm M_p=1.673 \times10^{-24}g$ to convert the central initial density from particles per volume unit to grams per volume unit. We apply Eq. \ref{eq:Pram} with the parameters in Table \ref{table:RPS_force} and find the following values for each cluster: $\rm P_{Virgo} = 2.1\times10^{-11} ~g~cm^{-1}~s^{-2}$, $\rm P_{Hydra} = 3.9\times10^{-12}~g~cm^{-1}~s^{-2}$ and $\rm P_{Fornax} = 6.2\times10^{-13}~g~cm^{-1}~s^{-2}$. We see that for this general scenario Virgo exhibits the strongest ram pressure, followed by Hydra I, and Fornax showing the weakest, with a difference of an order of magnitude between each.

\end{appendix}

\end{document}